\documentclass[aps,pra,twocolumn,superscriptaddress,10pt,notitlepage,nofootinbib]{revtex4-1}

\usepackage[pdftex]{graphicx}   
\usepackage{amsmath}
\usepackage{dcolumn}            

\usepackage{multirow}

\usepackage[table]{xcolor}

\usepackage{feynmp}             
\usepackage{color}              

\newcolumntype{f}[1]{D{.}{.}{#1}}

\newcommand\cnt[2]{\multicolumn{#1}{c}{#2}}
\newcommand\lft[2]{\multicolumn{#1}{l}{#2}}

\newcommand\cntl[2]{\multicolumn{#1}{c|}{#2}}
\newcommand\cntll[2]{\multicolumn{#1}{c||}{#2}}

\definecolor{darkgreen}{rgb}{0,0.7,0.2}

\newcommand{\mup}{\ensuremath{\mu}{\rm p}}
\newcommand{\mud}{\ensuremath{\mu}{\rm d}}
\newcommand{\muHet}{\ensuremath{\mu^3}{\rm He}}
\newcommand{\muHef}{\ensuremath{\mu^4}{\rm He}}
\newcommand{\rrd}{\ensuremath{\langle r_\mathrm{d}^2 \rangle}}
\newcommand{\rr}{\ensuremath{\langle r^2 \rangle}}

\newcommand{\rp}{\ensuremath{r_\mathrm{p}}}
\newcommand{\rd}{\ensuremath{r_\mathrm{d}}}

\newcommand{\etal}{{\it et\,al.}}

\newcommand{\ELSTPE}{\ensuremath{\Delta E_\mathrm{TPE}^\mathrm{LS}}}
\newcommand{\ELSFriar}{\ensuremath{\Delta E_\mathrm{Friar}^\mathrm{LS}}}

\newcommand{\ELSinelast}{\ensuremath{\Delta E_\mathrm{inelastic}^\mathrm{LS}}}

\newcommand{\EHFSTPE}{\ensuremath{\Delta E_\mathrm{TPE}^\mathrm{HFS}}}
\newcommand{\EHFSZemach}{\ensuremath{\Delta E_\mathrm{Zemach}^\mathrm{HFS}}}
\newcommand{\EHFSinelast}{\ensuremath{\Delta E_\mathrm{inelastic}^\mathrm{HFS}}}

\renewcommand{\baselinestretch}{1.2}

\def\LSVAL{\ensuremath{228.77356}}
\def\LSERR{\ensuremath{  0.00075}}
\def\TABLSVAL{{\ensuremath{\bf 228}}\bf .{\ensuremath{\bf 77356}}}
\def\TABLSERR{{\ensuremath{\bf   0}}\bf .{\ensuremath{\bf 00075}}}
\def\LSVALERR{\LSVAL \ensuremath{(75)}}

\def\POLVALRND{\ensuremath{1.7091}}
\def\POLERRRND{\ensuremath{0.0200}}  
\def\POLVALFINAL{\ensuremath{1.709}}
\def\POLERRFINAL{\ensuremath{0.020}}  
\def\POLVALOUR{\ensuremath{1.7091}}
\def\POLERROUR{\ensuremath{0.0146}} 
\def\TABPOLVAL{{\ensuremath{\bf 1}}\bf .{\ensuremath{\bf 7091}}}
\def\TABPOLERR{{\ensuremath{\bf 0}}\bf .{\ensuremath{\bf 0146}}}

\def\FSVAL{\ensuremath{8.86412}}

\def\TABFSVAL{{\ensuremath{\bf 8}}\bf .{\ensuremath{\bf 86412}}}
\def\TABFSERR{{\ensuremath{\bf 0}}\bf .{\ensuremath{\bf 00016}}}
\def\FSVALERR{\FSVAL \ensuremath{(16)}}

\def\HFSVAL{\ensuremath{6.39675}}

\def\TABHFSVAL{{\ensuremath{\bf 6}}\bf .{\ensuremath{\bf 39675}}}
\def\TABHFSERR{{\ensuremath{\bf 0}}\bf .{\ensuremath{\bf 00494}}}
\def\HFSVALERR{\HFSVAL \ensuremath{(494)}}

\def\MPQ{Max--Planck--Institut f{\"u}r Quantenoptik, 85748 Garching, Germany.}
\def\ETH{Institute for Particle Physics, ETH Zurich, 8093 Zurich, Switzerland.}
\def\PSI{Paul-Scherrer-Institute, 5232 Villigen, Switzerland.}

\begin{document}

\title{Theory of the \boldmath{$n=2$} levels in muonic deuterium}

\author{Julian J. Krauth}
\email{julian.krauth@mpq.mpg.de}
\affiliation{\MPQ}
\author{Marc~Diepold}
\affiliation{\MPQ}
\author{Beatrice~Franke}
\affiliation{\MPQ}
\author{Aldo~Antognini}
\affiliation{\ETH}
\affiliation{\PSI}
\author{Franz~Kottmann}
\affiliation{\ETH}
\author{Randolf~Pohl}
\email{randolf.pohl@mpq.mpg.de}
\affiliation{\MPQ}


\begin{abstract}
The present knowledge of Lamb shift, fine- and hyperfine structure of
the 2S and 2P states in muonic deuterium is reviewed in anticipation
of the results of a first measurement of several $\mathrm{2S-2P}$ transition 
frequencies in muonic deuterium (\mud).
A term-by-term comparison of all available sources reveals reliable values 
and uncertainties of the QED and nuclear structure-dependent contributions
to the Lamb shift, which are essential for a determination of the deuteron
rms charge radius from \mud{}.
Apparent discrepancies between different sources are resolved, 
in particular for the difficult two-photon exchange contributions.
Problematic single-sourced terms are identified which require independent 
recalculation.
%
\end{abstract}

\maketitle

\section{Introduction}
\label{sec:Intro}

Laser spectroscopy of
$\mathrm{2S \rightarrow 2P}$ Lamb shift transitions
in muonic atoms and
ions promises a tenfold improvement in our knowledge of charge and magnetic
radii of the lightest nuclei ($Z=1, 2$ and higher).
Our recent measurement~\cite{Pohl:2010:Nature_mup1,Antognini:2013:Science_mup2}
of the 2S Lamb shift and the 2S hyperfine splitting
(HFS) in muonic hydrogen, \mup{}, in combination with accurate theoretical
calculations by many authors, summarized in Ref.~\cite{Antognini:2013:Annals},
has revealed a proton root-mean-square (rms) charge radius of 
\begin{equation}
\label{eq:Rp_muH}
\rp ~ = ~ 0.84087\,(26)^{\rm exp}\,(29)^{\rm theo} ~\mathrm{fm} ~ = ~ 0.84087\,(39) \,\mathrm{fm}.
\end{equation}

This is an order of magnitude more accurate than the value of $\rp = 0.8775(51)$\,fm
evaluated in the CODATA least-squares adjustment~\cite{Mohr:2012:CODATA10} of
elastic electron-proton scattering~\cite{Sick:2003:RP,Bernauer:2010:NewMainz}
and many precision measurements in electronic hydrogen~\cite{Biraben:2009:SpectrAtHyd}.

Most strikingly, however, the two values differ by 7 combined standard
deviations ($7 \sigma$). Despite numerous attempts in recent years to explain
this ``proton radius puzzle'', it remains a
mystery~\cite{Pohl:2013:ARNPS,Carlson:2015:Puzzle}. Taken at face value, this
$7\sigma$ discrepancy constitutes one of the biggest discrepancies 
in the Standard Model.
Further data are clearly required to shed light on this puzzle.

Muonic deuterium, \mud{}, has been measured in the same beam time as 
\mup{}~\cite{Pohl:2010:Nature_mup1,Antognini:2013:Science_mup2}, and the
data are now nearing publication~\cite{CREMA:muD}. 
We anticipate here that the experimental accuracy 
of the various $\mathrm{2S-2P}$ transitions is of the order of 1\,GHz, or,
equivalently, $\sim 0.004$\,meV~\footnote{\protect{1\,meV $\widehat{=}$ 241.799\,GHz}}.
Ideally, theory should be accurate on the level of 0.001\,meV to exploit the
experimental precision, and to determine the deuteron charge radius, \rd{}, with
tenfold better accuracy, compared to the CODATA value~\cite{Mohr:2012:CODATA10}
\begin{equation}
\label{eq:Rd_CODATA}
\rd\,\mathrm{(CODATA)} = 2.1424(21) \,\mathrm{fm}.
\end{equation}
The CODATA value originates from a least-squares adjustment of 
a huge amount of input values, such as the deuteron charge radius from elastic
electron scattering~\cite{Sick:1996:RD,Sick:1998:RD}
\begin{equation}
\label{eq:Rd_scatt}
\rd\,\mathrm{(e-d~scatt.)} = 2.130(10) \,\mathrm{fm},
\end{equation}
but also the proton radius from electron scattering~\cite{Sick:2003:RP,Bernauer:2010:NewMainz}. These radii are connected because the CODATA adjustment includes many transition frequencies in hydrogen (H) and deuterium (D)~\cite{Biraben:2009:SpectrAtHyd,Mohr:2012:CODATA10}. In particular,
the squared deuteron-proton charge radius difference, 
\begin{equation}
\label{eq:iso_HD}
\rd^2 - \rp^2 = 3.82007\,(65) \,\mathrm{fm}^2
\end{equation}
is known with high precision from laser spectroscopy of
the isotope shift of the $\mathrm{1S-2S}$ transition in
electronic hydrogen and deuterium~\cite{Parthey:2010:PRL_IsoShift},
and state-of-the-art theory~\cite{Jentschura:2011:IsoShift}.
Using Eq.~(\ref{eq:iso_HD})
and the muonic hydrogen proton radius given in Eq.~(\ref{eq:Rp_muH})
we determined a value of~\cite{Antognini:2013:Science_mup2}
\begin{equation}
\label{eq:Rd_muH}
\rd\,\mathrm{(muonic~\rp)} = 2.12771\,(22) \,\mathrm{fm}.
\end{equation}
Note that the discrepancy of the deuteron charge radii given in 
Eq.~(\ref{eq:Rd_CODATA}) and 
Eq.~(\ref{eq:Rd_muH}) is not a new discrepancy, but rather a 
result of the proton radius discrepancy:
Both values of the deuteron radius depend on the isotope shift in Eq.~(\ref{eq:iso_HD}).
Hence, discrepant values of the proton radius will result in discrepant values
of the deuteron radius.

The upcoming \mud{} data~\cite{CREMA:muD}, on the other hand, will provide a
``muonic'' value of the deuteron radius that is {\em independent} of the 
proton charge radius. As such, it will shed new light on the proton radius puzzle.

We anticipate here that the theory of the Lamb shift in muonic deuterium 
is limiting the accuracy of the deuteron charge radius from \mud{}, mainly due
to the uncertainty of the deuteron polarizability contribution of 
\POLERRFINAL\,meV
which corresponds to a relative uncertainty of 1\%.
Nevertheless, the deuteron charge radius from \mud{}~\cite{CREMA:muD} will have a nearly three times
smaller uncertainty than the current CODATA value (Eq.~(\ref{eq:Rd_CODATA})).

To put this uncertainty of \POLERRFINAL\,meV into another perspective: The ``proton
radius puzzle'' in muonic hydrogen, when expressed as a ``missing part'' in
the theory of muonic hydrogen, amounts to 0.329\,meV.
%

This article is organized as follows: We first summarize the current knowledge
of the muonic deuterium Lamb shift theory (Sec.~\ref{sec:LS}) which is
required to determine the deuteron rms charge radius \rd{} 
from the \mud{} measurement~\cite{CREMA:muD}.
We separate the Lamb shift theory into ``radius-independent'' terms 
that do not depend on the nuclear structure 
(Sec.~\ref{sec:LS:QED}),
terms that depend explicitly on the rms charge radius (Sec.~\ref{sec:LS:Radius}),
and the deuteron polarizability contribution that constitutes the main 
theoretical limitation (Sec.~\ref{sec:LS:Pol}).
In Sec.~\ref{sec:HFS}, we list all contributions to the 2S hyperfine splitting 
(HFS) in \mud.
The 2S HFS depends on the magnetic properties of the deuteron through the
Zemach radius. Other nuclear structure contributions matter, too, so
we separate again terms:
Sec.~\ref{sec:HFS:QED} lists the terms that do not depend 
strongly on the deuteron structure,
Sec.~\ref{sec:HFS:Rz} is devoted to the Zemach correction,
Sec.~\ref{sec:HFS:Pol} is concerned with the deuteron polarizability contribution
which has recently been calculated for the first 
time~\cite{Faustov:2014:HFS_mud}. This term constitutes the main uncertainty
for the 2S HFS prediction.
Additional contributions to the 2S HFS are mentioned in Sec.~\ref{sec:HFS:other}.
The 2P fine structure is summarized in Sec.~\ref{sec:fs},
and the 2P fine- and hyperfine level structure, including level mixing,
is given in Sec.~\ref{sec:2PHFS}.

The sign convention in this article is such that the final, 
measured energy difference (Lamb shift, 2S-HFS, fine structure) is always
{\em positive}. For the fine and hyperfine splittings this convention
is the natural choice adopted by all other authors, too.
For the Lamb shift, however, some authors calculate 2S {\em level shifts} and
their published values have the opposite sign.
This is because the 2S level is lower (more bound) than the 2P level (due to
the dominant vacuum-polarization term item \#1 in Tab.~\ref{tab:LS:QED}),
see Fig.~\ref{fig:energy_level}, and
{\em positive} level shifts {\em decrease} the measured 2P-2S energy difference.
The numbers we quote are all matched to our sign convention.

Item numbers \# in the first column of Tab.~\ref{tab:LS:QED} and 
Tab.~\ref{tab:hfs} 
follow the nomenclature in Ref.~\cite{Antognini:2013:Annals}.
In the tables, we usually identify the ``source'' of all values
entering ``our choice'' by the first name of the (group of) authors
given in adjacent columns (e.g.\ ``B'' for Borie).
We denote as average ``avg.'' in the tables the center of the band covered by 
all values $v_i$ under consideration, 
with an uncertainty of half the spread, i.e.\
\begin{equation}
  \label{eq:avg}
  \begin{aligned}
  \mathrm{avg.} ~ = & ~ &
  \frac{1}{2}\,\big[ {\rm MAX}(v_i) + {\rm MIN}(v_i) \big] \\[1ex]
  & \pm &
  \frac{1}{2}\,\big[ {\rm MAX}(v_i) - {\rm MIN}(v_i) \big]
  \end{aligned}
\end{equation}

Throughout the paper, 
$Z$ denotes the nuclear charge with $Z=1$ for the deuteron,
$\alpha$ is the fine structure constant,
$m_r$ is the reduced mass of the muon-deuteron system.
``VP'' is short for ``vacuum polarization'',
``SE'' is ``self-energy'',
``RC'' is ``recoil correction''.
``Perturbation theory'' is abbreviated as ``PT'', and SOPT and TOPT denote
$2^{\rm nd}$ and $3^{\rm rd}$ order PT, respectively.

\section{Overview}
\label{sec:overview}

\begin{figure*}[t]
\centering
 \includegraphics[width=0.75\linewidth]{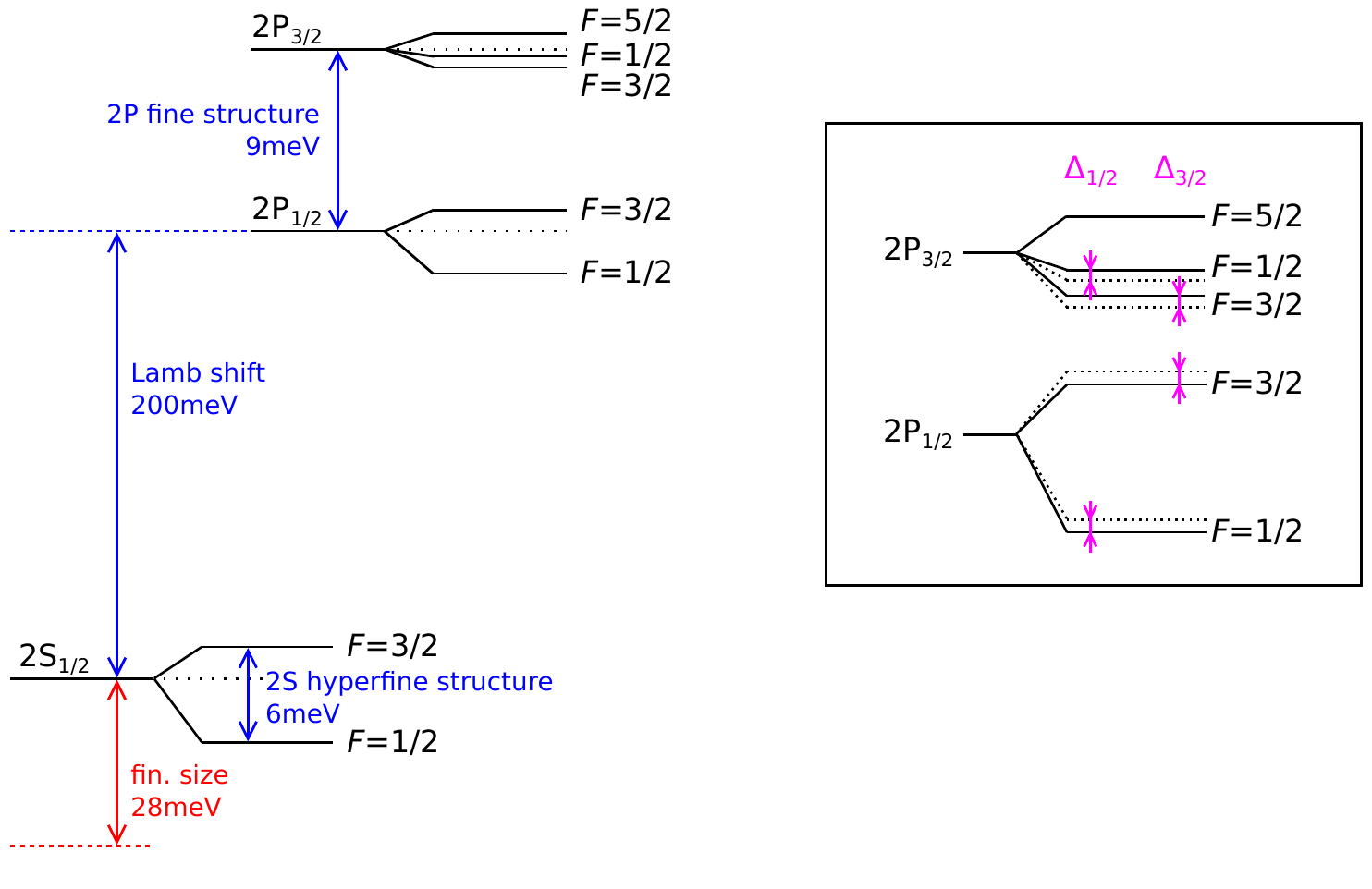}
 \caption{The 2S and 2P energy levels in muonic deuterium. 
  The inset on the right displays the shifts $\Delta_{1/2}$ and $\Delta_{3/2}$ 
  of the 2P$(F=1/2)$ and 2P$(F=3/2)$ levels
  due to the mixing of the $F=1/2$ and $F=3/2$ states, 
  respectively, as described in Sec.\,\ref{sec:2PHFS}.
  The figure is not to scale.}
  \label{fig:energy_level}
\end{figure*}

The $n=2$ levels in muonic deuterium are sketched in Fig.~\ref{fig:energy_level}.
The Lamb shift, i.e.\ the splitting between the 2S and the 2P$_{1/2}$ state,
is sensitive to the rms charge radius of the deuteron, as detailed in Sec.~\ref{sec:LS}.
In contrast, 2S hyperfine splitting (HFS) is caused by the magnetic interaction 
between the muon spin and the magnetic moment of the deuteron.
The finite deuteron size
results in a finite magnetization distribution inside the deuteron, and
makes the 2S HFS depend on the so-called Zemach radius of the deuteron,
as explained in Sec.~\ref{sec:HFS}.

The first calculation of the Lamb shift in muonic deuterium was published by
Carboni~\cite{Carboni:1973:LS_muD} in 1973. 
More elaborate calculations of QED effects in muonic atoms were
introduced with the seminal paper by Borie and Rinker~\cite{BorieRinker:1982:muAtoms} in 1982.

Later, Pachucki~\cite{Pachucki:1996:LSmup} and Borie~\cite{Borie:2005:LSmup} 
presented very detailed calculations of many terms for muonic hydrogen.
Borie then extended her \mup{} calculations~\cite{Borie:2005:LSmup}
to the case of muonic deuterium~\cite{Borie:2005:LSmud}.
After our measurements in muonic hydrogen~\cite{Pohl:2010:Nature_mup1,Antognini:2013:Science_mup2}
and deuterium, 
Borie revisited the theory of the $n=2$ energy levels in light muonic atoms 
(\mup, \mud, \muHet, and \muHef)
in Ref.~\cite{Borie:2012:LS_revisited_AoP}.
The published paper~\cite{Borie:2012:LS_revisited_AoP} 
(available online 6 Dec.\ 2011)
has subsequently been superseded by the arXiv version of the
paper~\cite{Borie:2014:arxiv_v7}. At the time of this writing, Borie's paper on the arXiv has 
reached version~7 (dated 21 Aug.\ 2014).
This is the first source of knowledge on \mud{} summarized in here.

The second source is the group around Faustov, Krutov, Martynenko \etal{},
termed ``Martynenko'' in here for simplicity. They have published an
impressive set of papers on theory of energy levels in light muonic atoms. 
At the time of this writing, the 2011 paper~\cite{Krutov:2011:PRA84_052514} 
was the most recent one on the Lamb shift in \mud{},
and we based our summary on this paper.
Later, Ref.~\cite{Martynenko:2014:muD_Theory} from the Martynenko group 
appeared, with only minor differences in the results
compared to Ref.~\cite{Krutov:2011:PRA84_052514}.
For simplicity, we still base our compilation of Lamb shift contributions 
on the earlier, more detailed, paper~\cite{Krutov:2011:PRA84_052514}.
In particular, equation numbers and table entries refer to 
Ref.~\cite{Krutov:2011:PRA84_052514}, unless otherwise noted.
For the 2S HFS, Ref.~\cite{Faustov:2014:HFS_mud} is the main source of
numbers from the Martynenko group.

After the advent of the proton radius puzzle from muonic hydrogen, many groups
have revisited and improved the theory on muonic hydrogen in an (unsuccessful)
attempt to identify wrong or missing theory terms large enough to solve the
puzzle (see our compilation~\cite{Antognini:2013:Annals} for a detailed
overview). Thankfully, two groups have \mbox{(re-)}calculated many terms not only
for the case 
of muonic hydrogen, but also for muonic deuterium (and \muHet{} and \muHef{}):
Jentschura, and Karshenboim's group with Ivanov, Korzinin, and Shelyuto,
have published many papers 
on muonic deuterium which are included in the present compilation.

\section{Lamb shift in muonic deuterium}
\label{sec:LS}

\subsection{QED contributions}
\label{sec:LS:QED}

\begin{figure}[]
  \begin{center}
    \begin{fmffile}{uehling}
      \begin{fmfgraph*}(80,80)
        \fmftop{i1,o1}
        \fmfbottom{i2,o2}
        \fmf{plain,tension=1.0}{i1,v1,o1}
        
        \fmf{plain,width=3}{i2,v2,o2}
        \fmf{phantom,tension=0.0001}{v1,c1,xx1,c2,v2}
        \fmf{photon,tension=0}{v1,c1}
        \fmf{photon,tension=0}{c2,v2}
        \fmf{plain,left,tension=0}{c1,c2,c1}
        \fmffreeze
        \fmfpoly{phantom}{c1,l1,c2,l2}
        \fmfv{label.angle=180,label=$e$}{l1}
        \fmfv{label=d}{i2}
        \fmfv{label.angle=-150,label=$\mu$}{i1}
      \end{fmfgraph*}
    \end{fmffile}
  \end{center}
  \caption{Item \#1, the leading order 1-loop electron vacuum polarization
    (eVP), also called Uehling term.}    
  \label{fig:uehling}
\end{figure}
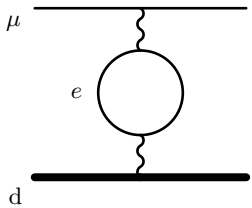

One-loop electron vacuum polarization (eVP) (Fig.~\ref{fig:uehling}),
the so-called Uehling term~\cite{Uehling:1935:VP},
accounts for 99.5\,\% of the nuclear-structure-independent part of the Lamb
 shift in
\mud{}. It is therefore mandatory to double-check this term as thoroughly as
possible.\\
%
%
%
Borie has argued~\cite{BorieRinker:1982:muAtoms,Borie:2005:LSmud,Borie:2012:LS_revisited_AoP,Borie:2014:arxiv_v7}
that the Uehling term should ideally not be treated perturbatively. Instead, the
Dirac equation should be solved numerically after adding the Uehling potential to
the electrostatic Coulomb potential.
For light muonic atoms such as muonic deuterium, however,
both approaches should give accurate results~\cite{Borie:2014:arxiv_v7}.
This has been demonstrated for muonic {\em hydrogen}, where 
the nonperturbative result of Indelicato~\cite{Indelicato:2012:Non_pert} is
in excellent agreement with the perturbative results of 
Pachucki~\cite{Pachucki:1996:LSmup} and Borie~\cite{Borie:2005:LSmup},
see Ref.~\cite{Antognini:2013:Annals}.

%
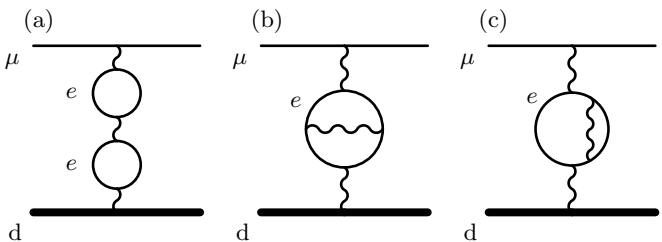
\begin{figure}[b!]
  \setlength{\unitlength}{0.0032\columnwidth}
  \begin{center}
    \begin{minipage}{0.3\columnwidth}
      (a)\hfill\mbox{~}\\
      \vspace{1ex}
      \begin{fmffile}{item_4}
        \begin{fmfgraph*}(80,80)
          \fmfstraight
          \fmftopn{t}{7}
          \fmfbottomn{b}{7}
          \fmf{plain,tension=1.0}{t1,t7}
          \fmf{plain,width=3        }{b1,b7}
          \fmf{phantom,tension=0.0001}{t4,c1,x1,c2,c4,x2,c5,b4}
          \fmf{photon,tension=0}{t4,c1}
          \fmf{photon,tension=0}{c2,c4}
          \fmf{photon,tension=0}{c5,b4}
          \fmf{plain,left,tension=0}{c1,c2,c1}
          \fmf{plain,left,tension=0}{c4,c5,c4}
          \fmffreeze
          \fmfv{label=d}{b1}
          \fmfv{label.angle=-150,label=$\mu$}{t1}
          \fmfpoly{phantom}{c1,l1,c2,l2}
          \fmfv{label.angle=180,label=$e$}{l1}
          \fmfpoly{phantom}{c4,l3,c5,l4}
          \fmfv{label.angle=180,label=$e$}{l3}
        \end{fmfgraph*}
      \end{fmffile}
    \end{minipage}
    \hfill
    \begin{minipage}{0.3\columnwidth}
      (b)\hfill\mbox{~}\\
      \vspace{1ex}
      \begin{fmffile}{item_4b}
        \begin{fmfgraph*}(80,80)
          \fmfstraight
          \fmftopn{t}{3}
          \fmfbottomn{b}{3}
          \fmf{plain}{t1,t2,t3}
          \fmf{plain,width=3}{b1,b2,b3}
          \fmf{photon}{t2,c0}
          \fmf{photon}{c4,b2}
          \fmfpoly{smooth,pull=?,tension=0.5}{c0,c1,c2,c3,c4,c5,c6,c7}
          \fmffreeze
          \fmf{photon}{c2,c6}
          \fmfv{label=d}{b1}
          \fmfv{label.angle=-150,label=$\mu$}{t1}
          \fmfv{label.angle=180,label=$e$}{c1}
        \end{fmfgraph*}
      \end{fmffile}
    \end{minipage}
    \hfill
    \begin{minipage}{0.3\columnwidth}
      (c)\hfill\mbox{~}\\
      \vspace{1ex}
      \begin{fmffile}{item_4c}
        \begin{fmfgraph*}(80,80)
          \fmfstraight
          \fmftopn{t}{3}
          \fmfbottomn{b}{3}
          \fmf{plain}{t1,t2,t3}
          \fmf{plain,width=3}{b1,b2,b3}
          \fmf{photon}{t2,c0}
          \fmf{photon}{c6,b2}
          \fmfpoly{smooth,pull=?,tension=0.8}{c0,c1,c2,c3,c4,c5,c6,c7,c8,c9,c10,c11}
          \fmffreeze
          \fmf{photon}{c7,c11}
          \fmfv{label=d}{b1}
          \fmfv{label.angle=-150,label=$\mu$}{t1}
          \fmfv{label.angle=180,label=$e$}{c1}
        \end{fmfgraph*}
      \end{fmffile}
    \end{minipage}
  \end{center}
  \caption{Item \#4, the two-loop eVP 
    (K\"allen-Sabry) contribution.
  }
  \label{fig:item_4}
\end{figure}

For muonic {\em deuterium}, only perturbative calculations exist,
albeit with two slightly different approaches:

Martynenko~\etal{}~\cite{Krutov:2011:PRA84_052514},
Jentschura~\cite{Jentschura:2011:PRA84_012505,Jentschura:2011:SemiAnalytic}
and Karshenboim~\etal{}~\cite{Karshenboim:2010:NRalpha5_mup,Karshenboim:2012:PRA85_032509} 
calculate the leading order eVP nonrelativistically 
(item~\#1 in Tab.~\ref{tab:LS:QED}),
and apply a relativistic correction (item~\#2).
The most important item~\#1 is in excellent agreement for all three authors,
as well as with the value of 227.635\,meV obtained 
by Carboni~\cite{Carboni:1973:LS_muD} in 1973.
For item~\#2 see below.\\
Borie, in contrast, uses relativistic Dirac wave functions to evaluate
the relativistic Uehling term (item~\#3).
The relativistic recoil
correction to eVP of order $\alpha(Z\alpha)^4$ (item~\#19) has to be added, to
be able to compare all four results.\\
It is very reassuring that these results are in excellent agreement,
with one exception: 
Item~\#2, {\em rel. corr. (Breit-Pauli)}, from 
Martynenko~\cite{Krutov:2011:PRA84_052514}, 0.0177\,meV, 
differs from the value of 0.02178\,meV, calculated by the other three groups,
who agree: Borie~\cite{Borie:2014:arxiv_v7} Tab.~1,
Jentschura~\cite{Jentschura:2011:relrecoil} Tab.~I and \cite{Jentschura:2011:SemiAnalytic} Eq.~(17), and
Karshenboim~\cite{Karshenboim:2012:PRA85_032509}~Tab.~IV.
This item \#2 should be the sum of Martynenko's
rows 7 and 10 ({\em relativistic and VP corrections of order $\alpha (Z
\alpha)^2$ in first and second order PT}).
For muonic helium-3 and -4 ions~\cite{Krutov:2014:JETP120_73}
the sum agrees exactly with the numbers given by
Jentschura in Ref.~\cite{Jentschura:2011:PRA84_012505}~Eq.~(17) and by 
Karshenboim in Ref.~\cite{Karshenboim:2012:PRA85_032509}~Tab.~IV.  
Martynenko confirmed that their value for our item~\#2 for muonic deuterium
(their rows 7 and 10 in Tab.~I of Ref.~\cite{Krutov:2011:PRA84_052514})
contains an error~\cite{Martynenko:PC:2015}.\\
The average of items \#1+\#2 or \#3+\#19 is thus calculated from the other
three sources only, with excellent agreement:
\begin{multline}
\label{eq:LS:eVP}
\Delta\mathrm{E~(one-loop~eVP~with~rel.~corr.)} \\
= 227.65658  \pm ~0.00020 \,\mathrm{meV}
\end{multline}

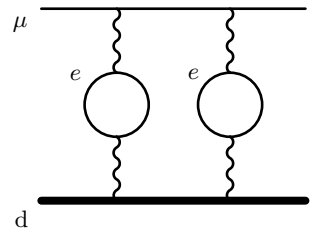
\begin{figure}[t]
  \setlength{\unitlength}{0.0037\columnwidth}
  \begin{center}
      \mbox{~}\vfill
      \begin{fmffile}{item_5}
        \begin{fmfgraph*}(110,80)
          \fmfstraight
          \fmftopn{t}{8}
          \fmfbottomn{b}{8}
          \fmf{plain,tension=1.0}{t1,t8}
          \fmf{plain,width=3        }{b1,b8}
          \fmf{phantom,tension=0.0001}{t3,c1,c2,b3}
          \fmf{photon,tension=0}{t3,c1}
          \fmf{photon,tension=0}{c2,b3}
          \fmf{plain,left,tension=0}{c1,c2,c1}
          \fmf{phantom,tension=0.0001}{t6,c3,c4,b6}
          \fmf{photon,tension=0}{t6,c3}
          \fmf{photon,tension=0}{c4,b6}
          \fmf{plain,left,tension=0}{c3,c4,c3}
          \fmffreeze
          \fmfv{label=d}{b1}
          \fmfv{label.angle=-150,label=$\mu$}{t1}
          \fmfpoly{phantom}{c1,l1,c2,l2}
          \fmfv{label.angle=110,label.dist=10,label=$e$}{l1}
          \fmfpoly{phantom}{c3,l3,c4,l4}
          \fmfv{label.angle=100,label.dist=10,label=$e$}{l3}
        \end{fmfgraph*}
      \end{fmffile}
  \end{center}
  \caption{Item \#5, the one-loop eVP in 2-Coulomb lines.}
  \label{fig:item_5}
\end{figure}

Our item~\#4, the {\em two-loop electron-VP correction}, 
usually called ``K\"allen-Sabry'' contribution~\cite{KallenSabry:1955},
displayed in Fig.~\ref{fig:item_4}, 
is the second largest ``purely QED'' contribution to the Lamb shift.
Borie~\cite{Borie:2014:arxiv_v7} and Martynenko~\cite{Krutov:2011:PRA84_052514}
give values which are in very good agreement.

Our item~\#5, the {\em one-loop eVP insertion in 2 Coulomb lines} shown in 
Fig.~\ref{fig:item_5}, has been calculated with very good agreement
by Borie~\cite{Borie:2014:arxiv_v7}, Martynenko~\cite{Krutov:2011:PRA84_052514},
and Jentschura~\cite{Jentschura:2011:SemiAnalytic}.

Karshenboim \etal{}~\cite{Karshenboim:2010:NRalpha5_mup} give the 
sum of items \#4 and \#5. It is in agreement with the sums from Borie and Martynenko.
We use Karshenboim's value because it is given with more significant digits.

\begin{figure}[b]
  \setlength{\unitlength}{0.0032\columnwidth}
  \begin{center}
    \mbox{~}
    \hfill
    \begin{minipage}{0.3\columnwidth}
      (a)\hfill\mbox{~}\vspace{1ex}
      \begin{fmffile}{lblone}
        \begin{fmfgraph*}(80,80)
          \fmfstraight
          \fmftop{i1,v1,o1}
          \fmfbottom{i2,v2,v3,v4,o2}
          \fmf{plain}{i1,v1,o1}
          \fmf{plain,width=3}{i2,v2,v3,v4,o2}

          \fmf{photon}{v1,c0}
          \fmf{photon}{c2,v3}

          \fmfpoly{default, pull=?, tension=0.3}{c0,c1,c2,c3}
          \fmffreeze

          \fmf{photon}{c1,v2}
          \fmf{photon}{c3,v4}
          \fmfv{label.angle=180,label=$e$}{c1}
          \fmfv{label=d}{i2}
          \fmfv{label.angle=-150,label=$\mu$}{i1}
        \end{fmfgraph*}
      \end{fmffile}
    \end{minipage}
    \hfill
    \begin{minipage}{0.3\columnwidth}
      (b)\hfill\mbox{~}\vspace{1ex}
      \begin{fmffile}{lbltwo}
        \begin{fmfgraph*}(80,80)
          \fmfstraight
          \fmftop{i1,v1,v2,o1}
          \fmfbottom{i2,v3,v4,o2}
          \fmf{plain}{i1,v1,v2,o1}
          \fmf{plain,width=3}{i2,v3,v4,o2}

          \fmf{photon}{v1,c0}
          \fmf{photon}{v2,c3}
          \fmf{photon}{v3,c1}
          \fmf{photon}{v4,c2}

          \fmfpoly{default, pull=?, tension=0.5}{c0,c1,c2,c3}

          \fmfv{label.angle=180,label=$e$}{c1}
          \fmfv{label=d}{i2}
          \fmfv{label.angle=-150,label=$\mu$}{i1}
        \end{fmfgraph*}
      \end{fmffile}
      \end{minipage}
    \hfill
    \begin{minipage}{0.3\columnwidth}
      (c)\hfill\mbox{~}\vspace{1ex}
      \begin{fmffile}{lblthree}
        \begin{fmfgraph*}(80,80)
          \fmfstraight
          \fmftop{i1,v1,v3,v4,o1}
          \fmfbottom{i2,v2,o2}
          \fmf{plain}{i1,v1,v3,v4,o1}
          \fmf{plain,width=3}{i2,v2,o2}

          \fmf{photon}{v3,c0}
          \fmf{photon}{c2,v2}

          \fmfpoly{default, pull=?, tension=0.3}{c0,c1,c2,c3}
          \fmffreeze

          \fmf{photon}{v1,c1}
          \fmf{photon}{v4,c3}
          \fmfv{label.angle=180,label=$e$}{c1}
          \fmfv{label=d}{i2}
          \fmfv{label.angle=-150,label=$\mu$}{i1}
        \end{fmfgraph*}
      \end{fmffile}
      \end{minipage}
  \end{center}
  \caption{The three contributions to Light-by-light scattering:
    (a) Wichmann-Kroll or ``1:3'' term, item~\#9,
    (b) Virtual Delbr\"uck or ``2:2'' term, item~\#10, and 
    (c) inverted Wichmann-Kroll or ``3:1'' term, item~\#9a$^\dagger$.}
  \label{fig:lbl}
\end{figure}
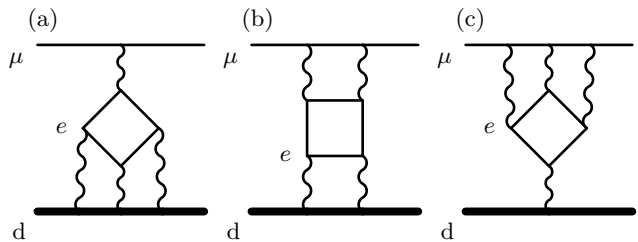

Light-by-light (LbL) scattering (see Fig.\,\ref{fig:lbl}) contains 3 terms,
 Wichmann-Kroll, or ``1:3'' LbL (item \#9 in \cite{Antognini:2013:Annals}),
 Virtual Delbr\"{u}ck, or ``2:2'' LbL (item \#10 in \cite{Antognini:2013:Annals}),
 and the ``inverted Wichmann-Kroll'' or ``3:1'' LbL
 (called ``new'' item in \cite{Antognini:2013:Annals}). 
For definiteness, we label this term ``item \#9a'' from now on.
Considerable cancellations occur in the sum of all three terms which
has been evaluated in Ref.~\cite{Karshenboim:2010:JETP_LBL}.
Both Borie and Martynenko calculate only \#9, and adopt the full result from 
Karshenboim~\cite{Karshenboim:2010:JETP_LBL}.
We use Karshenboim's result~\cite{Karshenboim:2010:JETP_LBL}.

\begin{figure}[b]
  \setlength{\unitlength}{0.0032\columnwidth}
  \begin{center}
      \mbox{~}
    \hfill
    \begin{minipage}{0.3\columnwidth}
      (a)\hfill\mbox{~}\\
      \vspace{5ex}
      \begin{fmffile}{item_11a}
        \begin{fmfgraph*}(80,80)
          \fmfstraight
          \fmftopn{t}{7}
          \fmfbottomn{b}{7}
          \fmf{plain,tension=1.0}{t1,t7}
          \fmf{plain,width=3        }{b1,b7}
          \fmf{photon,tension=0.1,left=1.0}{t3,t5}
          \fmf{phantom,tension=0.0001}{t4,c1,c2,b4}
          \fmf{photon,tension=0}{t4,c1}
          \fmf{photon,tension=0}{c2,b4}
          \fmf{plain,left,tension=0}{c1,c2,c1}
          \fmffreeze
          \fmfpoly{phantom}{c1,l1,c2,l2}
          \fmfv{label.angle=180,label=$e$}{l1}
          \fmfv{label=d}{b1}
          \fmfv{label.angle=-150,label=$\mu$}{t1}
        \end{fmfgraph*}
      \end{fmffile}
    \end{minipage}
    \hfill
    \begin{minipage}{0.3\columnwidth}
      (b)\hfill\mbox{~}\vspace{5ex}
      \begin{fmffile}{item_11b}
        \begin{fmfgraph*}(80,80)
          \fmfstraight
          \fmftopn{t}{7}
          \fmfbottomn{b}{7}
          \fmf{plain,tension=1.0}{t1,t7}
          \fmf{plain,width=3        }{b1,b7}
          \fmf{photon,tension=0.1,left=1.0}{t2,t4}
          \fmf{phantom,tension=0.0001}{t5,c1,c2,b5}
          \fmf{photon,tension=0}{t5,c1}
          \fmf{photon,tension=0}{c2,b5}
          \fmf{plain,left,tension=0}{c1,c2,c1}
          \fmffreeze
          \fmfpoly{phantom}{c1,l1,c2,l2}
          \fmfv{label.angle=180,label=$e$}{l1}
          \fmfv{label=d}{b1}
          \fmfv{label.angle=-150,label=$\mu$}{t1}
       \end{fmfgraph*}
      \end{fmffile}
    \end{minipage}
    \hfill
    \begin{minipage}{0.3\columnwidth}
      (c)\hfill\mbox{~}\vspace{5ex}
      \begin{fmffile}{item_11c}
        \begin{fmfgraph*}(80,80)
          \fmfstraight
          \fmftopn{t}{7}
          \fmfbottomn{b}{7}
          \fmf{plain,tension=1.0}{t1,t7}
          \fmf{plain,width=3        }{b1,b7}
          \fmf{photon,tension=0.1,left=1.0}{t4,t6}
          \fmf{phantom,tension=0.0001}{t3,c1,c2,b3}
          \fmf{photon,tension=0}{t3,c1}
          \fmf{photon,tension=0}{c2,b3}
          \fmf{plain,left,tension=0}{c1,c2,c1}
          \fmffreeze
          \fmfpoly{phantom}{c1,l1,c2,l2}
          \fmfv{label.angle=180,label=$e$}{l1}
          \fmfv{label=d}{b1}
          \fmfv{label.angle=-150,label=$\mu$}{t1}
        \end{fmfgraph*}
      \end{fmffile}
    \end{minipage}
  \end{center}
  \caption{Item~\#11, muon self-energy corrections to the electron vacuum polarization
    $\alpha^2 (Z\alpha)^4$.
    This figure is Fig.~2 from Jentschura~\cite{Jentschura:2011:AnnPhys1}.
    It corresponds to Fig.~6(a) from Karshenboim~\cite{Korzinin:2013:PRD88_125019}.}
  \label{fig:item_11}
\end{figure}
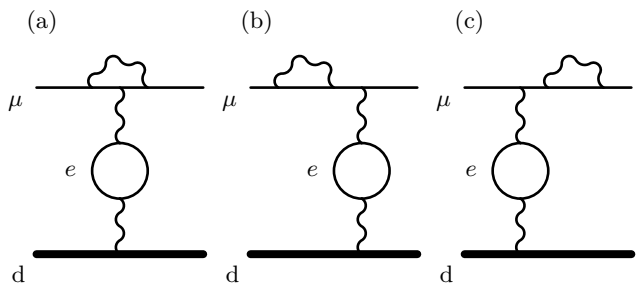

Item \#11 {\em muon self-energy (SE) correction to electron vacuum polarization (eVP) $\alpha^2 (Z\alpha)^4$} is
displayed in Fig.~\ref{fig:item_11}. 
Jentschura~\cite{Jentschura:2011:SemiAnalytic} (Eq.~(29b), Fig.\,2) and 
Karshenboim~\cite{Korzinin:2013:PRD88_125019} (Tab.VIII~(a), Fig.\,6a)
agree in the result for the complete calculation, -0.00306\,meV.
Martynenko~\cite{Krutov:2011:PRA84_052514}~Eq.~(80) calculates only the
contribution from Fig.~\ref{fig:item_11}(a), following 
Pachucki's Eq.~(39) in Ref.~\cite{Pachucki:1996:LSmup}.
Also Borie~\cite{Borie:2014:arxiv_v7} calculates part of this term in her 
Appendix~C.

Higher order corrections to the muon self-energy and vacuum polarization
are denoted items \#12, \#13, \#21, \#30$^*$ and \#31$^*$ in 
Tab.~\ref{tab:LS:QED}.
For muonic hydrogen~\cite{Antognini:2013:Annals} we used Borie's value of 
item \#21, noting that this includes item \#12.
Afterwards, Karshenboim~\etal{}~\cite{Korzinin:2013:PRD88_125019} have 
recalculated many of these small terms. 
We construct the corresponding sum from each source, which we average.

\begin{figure}[b]
  \setlength{\unitlength}{0.0032\columnwidth}
  \begin{center}
    %
    %
    \begin{minipage}{0.30\columnwidth}
      \mbox{~}\vfill
      \centering
      \begin{fmffile}{Mart_fig11b}
        \begin{fmfgraph*}(80,80)
          \fmfstraight
          \fmfleftn{i}{9}
          \fmfrightn{o}{9}
          \fmf{plain,tension=1.0}{i5,t1,t2,t3,o5}
          \fmf{plain,width=3}{i1,b2,o1}
          \fmf{photon,tension=0}{t2,b2}
          \fmf{phantom}{i7,c1}
          \fmf{phantom}{o7,c2}
          \fmf{photon,tension=0,left=0.3}{t1,c1}
          \fmf{photon,tension=0,left=0.3}{c2,t3}
          \fmf{plain,left,tension=0.5}{c1,c2,c1}
          \fmffreeze
          \fmfv{label.angle=100,label.dist=10,label=$e$}{c1}
          \fmfv{label=d}{i1}
          \fmfv{label.angle=-150,label=$\mu$}{i5}
        \end{fmfgraph*}
      \end{fmffile}
    \end{minipage}
  \end{center}
  \caption{Item \#12, {\em eVP loop in SE} are radiative corrections with VP effects.
    This is Fig.~11 from Martynenko~\cite{Krutov:2011:PRA84_052514} which
  is the same as Fig.~4 in Pachucki~\cite{Pachucki:1996:LSmup}.
  It is Karshenboim's Fig.~6(d) in Ref.~\cite{Korzinin:2013:PRD88_125019}.}
  \label{fig:Mart_Fig11}
  \label{fig:item_12}
\end{figure}
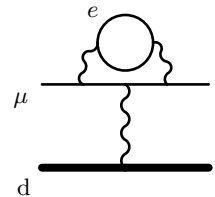

Item \#12 is shown in Fig.~\ref{fig:item_12}.
It is Martynenko's item 29.
This contribution has been confirmed by 
Karshenboim~\cite{Korzinin:2013:PRD88_125019}~Tab.~VIII~(d).
As mentioned in Ref.~\cite{Antognini:2013:Annals}, 
item~\#12 is included in Borie's value for item~\#21.

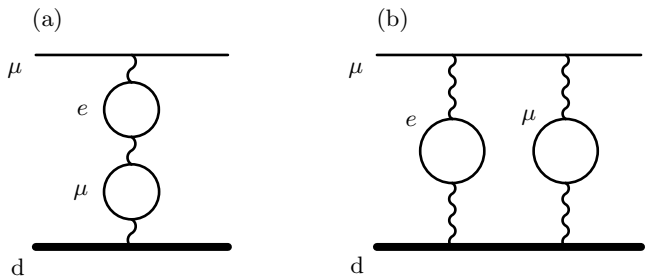
\begin{figure}[t]
  \setlength{\unitlength}{0.0037\columnwidth}
  \begin{center}
    \mbox{~ ~}
    \begin{minipage}{0.32\columnwidth}
      (a)\hfill\mbox{~}\\
      \vspace{2ex}
      \begin{fmffile}{item_13a}
        \begin{fmfgraph*}(80,80)
          \fmfstraight
          \fmftopn{t}{7}
          \fmfbottomn{b}{7}
          \fmf{plain,tension=1.0}{t1,t7}
          \fmf{plain,width=3        }{b1,b7}
          \fmf{phantom,tension=0.0001}{t4,c1,x1,c2,c4,x2,c5,b4}
          \fmf{photon,tension=0}{t4,c1}
          \fmf{photon,tension=0}{c2,c4}
          \fmf{photon,tension=0}{c5,b4}
          \fmf{plain,left,tension=0}{c1,c2,c1}
          \fmf{plain,left,tension=0}{c4,c5,c4}
          \fmffreeze
          \fmfv{label=d}{b1}
          \fmfv{label.angle=-150,label=$\mu$}{t1}
          \fmfpoly{phantom}{c1,l1,c2,l2}
          \fmfv{label.angle=180,label=$e$}{l1}
          \fmfpoly{phantom}{c4,l3,c5,l4}
          \fmfv{label.angle=180,label=$\mu$}{l3}
        \end{fmfgraph*}
      \end{fmffile}
    \end{minipage}
    \hfill
    \begin{minipage}{0.42\columnwidth}
      (b)\hfill\mbox{~}\\
      \vspace{2ex}
      \begin{fmffile}{item_13b}
        \begin{fmfgraph*}(110,80)
          \fmfstraight
          \fmftopn{t}{8}
          \fmfbottomn{b}{8}
          \fmf{plain,tension=1.0}{t1,t8}
          \fmf{plain,width=3        }{b1,b8}
          \fmf{phantom,tension=0.0001}{t3,c1,c2,b3}
          \fmf{photon,tension=0}{t3,c1}
          \fmf{photon,tension=0}{c2,b3}
          \fmf{plain,left,tension=0}{c1,c2,c1}
          \fmf{phantom,tension=0.0001}{t6,c3,c4,b6}
          \fmf{photon,tension=0}{t6,c3}
          \fmf{photon,tension=0}{c4,b6}
          \fmf{plain,left,tension=0}{c3,c4,c3}
          \fmffreeze
          \fmfv{label=d}{b1}
          \fmfv{label.angle=-150,label=$\mu$}{t1}
          \fmfpoly{phantom}{c1,l1,c2,l2}
          \fmfv{label.angle=110,label.dist=10,label=$e$}{l1}
          \fmfpoly{phantom}{c3,l3,c4,l4}
          \fmfv{label.angle=100,label.dist=10,label=$\mu$}{l3}
        \end{fmfgraph*}
      \end{fmffile}
    \end{minipage}
  \end{center}
  \caption{Item \#13, the mixed eVP-$\mu$VP contribution.}    
  \label{fig:item_13}
\end{figure}

Item \#13, {\em mixed muon-electron VP} is depicted in Fig.~\ref{fig:item_13}.
Borie and Martynenko calculated only the contribution from Fig.~\ref{fig:item_13}(a),
see Fig.~3 in Ref.~\cite{Borie:1974:HelvPhysAct}.
This is Martynenko's item 3, ``VP and MVP contribution in one-photon interaction''.
Karshenboim gives the sum of both diagrams in Fig.~\ref{fig:item_13} in 
Ref.~\cite{Korzinin:2013:PRD88_125019}~Tab.~VIII~(d).\\
%
%

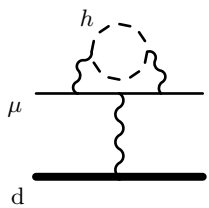
\begin{figure}[b]
  \setlength{\unitlength}{0.0032\columnwidth}
  \begin{center}
    \begin{minipage}{0.30\columnwidth}
      \mbox{~}\vfill
      \centering
      \begin{fmffile}{item_30}
        \begin{fmfgraph*}(80,80)
          \fmfstraight
          \fmfleftn{i}{9}
          \fmfrightn{o}{9}
          \fmf{plain,tension=1.0}{i5,t1,t2,t3,o5}
          \fmf{plain,width=3}{i1,b2,o1}
          \fmf{photon,tension=0}{t2,b2}
          \fmf{phantom}{i7,c1}
          \fmf{phantom}{o7,c2}
          \fmf{photon,tension=0,left=0.3}{t1,c1}
          \fmf{photon,tension=0,left=0.3}{c2,t3}
          \fmf{dashes,left,tension=0.5}{c1,c2,c1}
          \fmffreeze
          \fmfv{label.angle=100,label.dist=10,label=$h$}{c1}
          \fmfv{label=d}{i1}
          \fmfv{label.angle=-150,label=$\mu$}{i5}
        \end{fmfgraph*}
      \end{fmffile}
    \end{minipage}
  \end{center}
  \caption{Item \#30$^*$, hadronic VP in SE contribution,
  corresponds to Fig.~6(e) in Karshenboim~\cite{Korzinin:2013:PRD88_125019}.}
    \label{fig:item_30}
\end{figure}

Item \#30$^*$ (\#31$^*$) is somewhat similar to item~\#12 (\#13), 
with the electron (muon)  loop replaced by a
hadronic VP loop, see Fig.~(\ref{fig:item_30}).
It has only been calculated by Karshenboim~\cite{Korzinin:2013:PRD88_125019},
Tab.~VIII item (e) ((c)).

Item \#32, the muon VP in SE correction shown in in Fig.~\ref{fig:item_32},
is not included as a separate item in our Tab.~\ref{tab:LS:QED}.
It should already be automatically included in
any QED value which has been rescaled from the QED of electronic deuterium 
by a simple mass replacement $m_e \rightarrow m_\mu$~\cite{Karshenboim:PC:2015}.
The size of this item \#32 can be estimated from
the relationship found by Borie~\cite{Borie:1981:HVP}, 
that the ratio of hadronic to muonic VP is 0.66.
With Karshenboim's value of item~\#30$^*$ \cite{Korzinin:2013:PRD88_125019}
one would get
$\Delta E(\#32) = -0.000024 / 0.66 ~\mathrm{meV} = -0.000036 ~\mathrm{meV}$.

\begin{figure}[t]
  \setlength{\unitlength}{0.0037\columnwidth}
  \begin{center}
    \mbox{~ ~}
    \begin{minipage}{0.32\columnwidth}
      (a)\hfill\mbox{~}\\
      \vspace{2ex}
      \begin{fmffile}{item_31a}
        \begin{fmfgraph*}(80,80)
          \fmfstraight
          \fmftopn{t}{7}
          \fmfbottomn{b}{7}
          \fmf{plain,tension=1.0}{t1,t7}
          \fmf{plain,width=3        }{b1,b7}
          \fmf{phantom,tension=0.0001}{t4,c1,x1,c2,c4,x2,c5,b4}
          \fmf{photon,tension=0}{t4,c1}
          \fmf{photon,tension=0}{c2,c4}
          \fmf{photon,tension=0}{c5,b4}
          \fmf{plain,left,tension=0}{c1,c2,c1}
          \fmf{dashes,left,tension=0}{c4,c5,c4}
          \fmffreeze
          \fmfv{label=d}{b1}
          \fmfv{label.angle=-150,label=$\mu$}{t1}
          \fmfpoly{phantom}{c1,l1,c2,l2}
          \fmfv{label.angle=180,label=$e$}{l1}
          \fmfpoly{phantom}{c4,l3,c5,l4}
          \fmfv{label.angle=180,label=$h$}{l3}
        \end{fmfgraph*}
      \end{fmffile}
    \end{minipage}
    \hfill
    \begin{minipage}{0.42\columnwidth}
      (b)\hfill\mbox{~}\\
      \vspace{2ex}
      \begin{fmffile}{item_31b}
        \begin{fmfgraph*}(110,80)
          \fmfstraight
          \fmftopn{t}{8}
          \fmfbottomn{b}{8}
          \fmf{plain,tension=1.0}{t1,t8}
          \fmf{plain,width=3        }{b1,b8}
          \fmf{phantom,tension=0.0001}{t3,c1,c2,b3}
          \fmf{photon,tension=0}{t3,c1}
          \fmf{photon,tension=0}{c2,b3}
          \fmf{plain,left,tension=0}{c1,c2,c1}
          \fmf{phantom,tension=0.0001}{t6,c3,c4,b6}
          \fmf{photon,tension=0}{t6,c3}
          \fmf{photon,tension=0}{c4,b6}
          \fmf{dashes,left,tension=0}{c3,c4,c3}
          \fmffreeze
          \fmfv{label=d}{b1}
          \fmfv{label.angle=-150,label=$\mu$}{t1}
          \fmfpoly{phantom}{c1,l1,c2,l2}
          \fmfv{label.angle=110,label.dist=10,label=$e$}{l1}
          \fmfpoly{phantom}{c3,l3,c4,l4}
          \fmfv{label.angle=100,label.dist=10,label=$h$}{l3}
        \end{fmfgraph*}
      \end{fmffile}
    \end{minipage}
  \end{center}
  \caption{Item \#31$^*$, the mixed eVP- and hadronic VP contribution,
    comes from the Uehling correction to the hadronic VP correction.
    See Fig.~6(c) in Karshenboim~\cite{Korzinin:2013:PRD88_125019}.}
    \label{fig:item_31}
\end{figure}
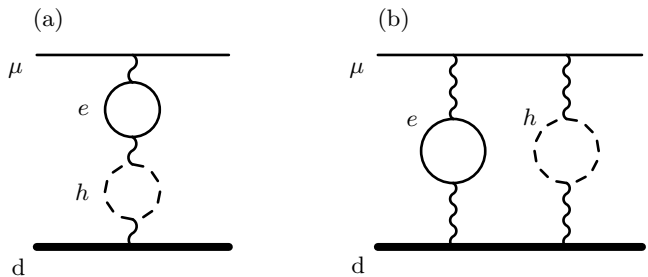

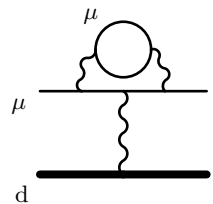
\begin{figure}[b]
  \setlength{\unitlength}{0.0032\columnwidth}
  \begin{center}
    \begin{minipage}{0.30\columnwidth}
      \mbox{~}\vfill
      \centering
      \begin{fmffile}{item_32}
        \begin{fmfgraph*}(80,80)
          \fmfstraight
          \fmfleftn{i}{9}
          \fmfrightn{o}{9}
          \fmf{plain,tension=1.0}{i5,t1,t2,t3,o5}
          \fmf{plain,width=3}{i1,b2,o1}
          \fmf{photon,tension=0}{t2,b2}
          \fmf{phantom}{i7,c1}
          \fmf{phantom}{o7,c2}
          \fmf{photon,tension=0,left=0.3}{t1,c1}
          \fmf{photon,tension=0,left=0.3}{c2,t3}
          \fmf{plain,left,tension=0.5}{c1,c2,c1}
          \fmffreeze
          \fmfv{label.angle=100,label.dist=10,label=$\mu$}{c1}
          \fmfv{label=d}{i1}
          \fmfv{label.angle=-150,label=$\mu$}{i5}
        \end{fmfgraph*}
      \end{fmffile}
    \end{minipage}
  \end{center}
  \caption{Item \#32, muon VP in SE contribution, is automatically included
  in a rescaled electronic deuterium QED value of higher order SE contributions
  (see text).}
  \label{fig:item_32}
\end{figure}

Item \#21, {\em higher order correction to $\mu$SE and $\mu$VP}, 
is Borie's {\em muon Lamb shift, higher orders}, calculated in her Appendix~C 
of Ref.~\cite{Borie:2014:arxiv_v7}.
This item includes item~\#12, 
{\em eVP loop in self-energy $\alpha^2 (Z\alpha)^4$},
as explained on p.~131 of Ref.~\cite{Antognini:2013:Annals}~\footnote{
There is a typo in footnote~f of Tab.~1 in Ref.~\cite{Antognini:2013:Annals},
where we wrote, item \#12 ``is part of \#22'' (instead of \#21).}.

The sum of items \#12, \#13, \#21, \#30$^*$\,\footnote{The asterisk~$^*$ 
indicates that this item had not been considered 
for muonic hydrogen in Ref.~\cite{Antognini:2013:Annals}.},
and \#31$^*$ agree
well enough to justify 
taking the average, $-0.00178 \pm 0.00014$\,meV as our choice.

Item \#14, {\em hadronic VP}, is evaluated by Borie~\cite{Borie:2014:arxiv_v7} (p.~5) as
0.013\,meV, who assigns a 5\% uncertainty to this estimate which is based on
Refs.~\cite{Borie:1981:HVP,Friar:1999:PRA59,BorieRinker:1982:muAtoms,Eides:2001:PhysRep}.\\
Martynenko's value of 0.0129\,meV \cite{Krutov:2011:PRA84_052514}, Tab.~1, row 31, 
agrees very nicely. They quote Ref.~\cite{Friar:1999:PRA59,Faustov:1999:HVP_muH}, and 
estimate the uncertainty to 5\% as well.\\
The previous items \#15 and \#16 in Ref.~\cite{Antognini:2013:Annals} 
are higher order corrections to the hadronic VP and
have only been calculated for muonic hydrogen by
Martynenko's group~\cite{Martynenko:2000,Martynenko:2001}, where they are
small (0.000047\,meV and -0.000015\,meV, respectively).  It is expected that
their magnitude will be similar in muonic deuterium.  Hence these terms are
included in the 5\% uncertainty assigned to item \#14.
%

Item~\#18 in Ref.~\cite{Antognini:2013:Annals} is the 
{\em recoil finite size} contribution in 
Borie~\cite{Borie:2012:LS_revisited_AoP,Borie:2014:arxiv_v7}.
According to Pachucki, this item~\#18, which was first calculated by 
Friar~\cite{Friar:1978:Annals}, 
should be discarded~\cite{Pachucki:PC:2015}, as we did for muonic hydrogen 
in Ref.~\cite{Antognini:2013:Annals}.

Item \#17 is the Barker-Glover correction~\cite{Barker:1955}, termed 
{\em additional recoil} by Borie.
This item includes the Darwin-Foldy (DF) term 
that arises from the Zitterbewegung of the nucleus. 
For a spin-1 nucleus such as the deuteron (as well as for the spin-0 $^4$He nucleus)
this DF term is absent~\cite{PachuckiKarshenboim:1995}.
Different conventions are used in the 
literature~\cite{Friar:1997:PRA56_4579,Khriplovich:1996:DF,Jentschura:2011:DF} 
which has caused some confusion (see Appendix~\ref{app:DF}).
As in the case of muonic hydrogen~\cite{Antognini:2013:Annals} 
(where the DF term is nonzero), we follow the 
``atomic physics'' convention~\cite{Jentschura:2011:DF},
also adopted by CODATA-2010~\cite{Mohr:2012:CODATA10}.
In this way, the charge radii from
muonic hydrogen~\cite{Pohl:2010:Nature_mup1,Antognini:2013:Science_mup2} and deuterium~\cite{CREMA:muD}
are directly comparable to the CODATA-2010 values~\cite{Mohr:2012:CODATA10},
as well as the electronic H-D-isotope shift given in Eq.~(\ref{eq:iso_HD}), 
all of which follow the same convention~\cite{Parthey:2010:PRL_IsoShift,Jentschura:2011:IsoShift}.

Item \#28$^\dagger$ is the {\em rad.\ (only eVP) RC $\alpha(Z\alpha)^5$} 
labeled ``new'' in Ref.~\cite{Antognini:2013:Annals}. For definiteness,
we enumerate it as item~\#28$^\dagger$.
It is the sum of three individual parts which sum up
to 0.000093\,meV in Ref.~\cite{Jentschura:2011:SemiAnalytic}, Eq.~(46b).\\
Martynenko's row 26 of Tab.~I in Ref.~\cite{Krutov:2011:PRA84_052514}, 
{\em recoil corr.\ to VP of order $\alpha(Z\alpha)^5$ (seagull term)}
is only the seagull term from the three terms evaluated by Jentschura, 
taken from Ref.~\cite{Jentschura:2011:PRA84_012505}, Eq.~(29).
We take the full result from
Ref.~\cite{Jentschura:2011:SemiAnalytic}, Eq.~(46b).

Item \#24 are radiative recoil corrections of order $\alpha (Z\alpha)^5$ and 
$(Z^2\alpha)(Z\alpha)^4$, first introduced for \mup{} by 
Pachucki~\cite{Pachucki:1996:LSmup}, Eq.~(51), based on 
Ref.~\cite{Pachucki:1995:RREC}.
Borie writes (p.~9 of Ref.~\cite{Borie:2014:arxiv_v7}) that
these terms correspond to Tab.~9 and two additional terms from Tab.~8
of the review of Eides \etal~\cite{Eides:2001:PhysRep}.
The sum is -0.00302\,meV for \mud{}.
Martynenko \etal{} (Ref.~\cite{Krutov:2011:PRA84_052514}, 
row 27 of Tab.~1, and Eq.~(71)),
on the other hand, evaluate only the terms from Tab.~9 of 
Eides~\cite{Eides:2001:PhysRep}, which gives -0.0026\,meV.
We use Borie's complete result.

Item \#29$^*$, the $\alpha^2(Z \alpha)^4 \,m$ contribution to the Lamb
shift, is new and has not been considered for muonic hydrogen
in Ref.~\cite{Antognini:2013:Annals}. We keep enumerating the
items, making this item~\#29$^*$~\footnotemark[3].\\
Martynenko gives this as the sum of rows 8 and 11 in Tab.~1 of
\cite{Krutov:2011:PRA84_052514}, 
{\em relativistic and two-loop VP corrections of order $\alpha^2(Z \alpha)^4$ in
1st and 2nd order PT}. They sum up to $-0.0002 + 0.0004 = 0.0002$\,meV.
Karshenboim calculated the complete correction of this order,
with recoil corrections included and calls it {\em eVP2} in Tab.~VIII 
of Ref.~\cite{Korzinin:2013:PRD88_125019}. Their value of 0.000203\,meV 
replaces Martynenko's rows 8 and 11~\cite{Karshenboim:PC:2015}.\\
Numerically this item \#29$^*$ is of little practical importance,
because it is so tiny: 0.000173\,meV in \mup{}, and 0.000203\,meV in \mud{}.
Of course, the calculation of this term was an important confirmation 
that previously uncalculated higher order terms are not responsible for the 
proton radius discrepancy.
It is very reassuring that the two different approaches of 
Martynenko~\cite{Krutov:2011:PRA84_052514} and 
Karshenboim~\cite{Korzinin:2013:PRD88_125019} give the same result.
Interestingly, there is no such an agreement for the cases 
of muonic helium-3 and -4 ions. In view of our recent measurements
in muonic helium-3 and -4~\cite{Antognini:2011:Conf:PSAS2010}, 
this disagreement may deserve further study, even though the size of the
terms is small compared to the overall uncertainty.


The sum of all contributions without explicit nuclear structure dependence
summarized in Tab.~\ref{tab:LS:QED} amounts to
\begin{equation}
  \label{eq:LS:QED}
  \Delta E^\mathrm{LS}_\mathrm{rad.-indep.} = \LSVAL ~ \pm ~ \LSERR \quad \mathrm{meV}.
\end{equation}

\clearpage
\begin{turnpage}
\begin{table}[h]
\renewcommand{\baselinestretch}{1.1}
\renewcommand{\arraystretch}{1.5}
\mbox{~}\vspace{-17ex}
\caption{All known radius-{\bf independent} contributions to the
  Lamb shift in \mud{}. Values are in meV.
  Item numbers ``\#'' in the 1st column follow the nomenclature of
  Ref.~\cite{Antognini:2013:Annals}, which in turn followed the Supplement 
  of Ref.~\cite{Pohl:2010:Nature_mup1}.
  Items ``\# `` with a dagger $^\dagger$ were labeled ``New'' in Ref.~\cite{Antognini:2013:Annals},
  but we introduce numbers here for definiteness.
  Items \# with an asterisk $^*$ denote new contributions in this compilation.\newline
  For Borie~\cite{Borie:2014:arxiv_v7} we refer to the most recent 
  arXiv version-7 (dated 21 Aug.\ 2014) which contains several corrections
  to the published paper~\cite{Borie:2012:LS_revisited_AoP} 
  (available online 6 Dec.\ 2011).
  For Martynenko, numbers \#1 to \#31 refer to rows in Tab.~I of 
  Ref.~\cite{Krutov:2011:PRA84_052514}.
  The values in their more recent paper~\cite{Martynenko:2014:muD_Theory} 
  agree exactly with the earlier values~\cite{Krutov:2011:PRA84_052514}.
  Numbers in parentheses refer to equations in the respective paper.
  }
\label{tab:LS:QED}
\setlength\tabcolsep{1mm}
\setlength{\extrarowheight}{0.2mm}
\begin{center}
    \fontsize{7pt}{7pt}\selectfont
    \vspace{-5ex}
    \hspace*{-12mm}
\begin{tabular}{ c | l |f{7} l | f{7} l | f{6} l | f{6}@{\qquad}l | f{5}@{ }f{5}  c  c}
      \hline
      \hline
\# & Contribution                                            & \cntl{2}{Borie}
                                                                              & \cntl{2}{Martynenko}
                                                                                                 & \cntl{2}{Jentschura}
                                                                                                                                        & \cntl{2}{Karshenboim}
                                                                                                                                             & \cnt{3}{Our choice} \\
   &                                                         & \cntl{2}{\cite{Borie:2014:arxiv_v7}}
                                                                          & \cntl{2}{~\cite{Krutov:2011:PRA84_052514}}  &  &  & & & \cnt{2}{value} & source & Fig. \\
\end{tabular}
\end{center}
  \end{table}
\end{turnpage}
\clearpage

%

\subsection{Radius-dependent contributions to the Lamb shift}
\label{sec:LS:Radius}
The radius-dependent contributions to the 
Lamb shift~\cite{Martynenko:2014:muD_Theory,Borie:2014:arxiv_v7, Karshenboim:2012:PRA85_032509}
are listed in Tab.~\ref{tab:LS:Radius}.
Generally, the finite size of the nucleus affects mainly the S states
whose wave function is nonzero at the origin, where the nucleus resides.
The main finite size contributions to the $n$S states
have been given to order $(Z\alpha)^6$ by Friar~\cite{Friar:1978:Annals}
\begin{widetext}
\begin{equation}
\label{eq:LS:FinSize}
\Delta E_{\rm fin.~size} = \frac{2\pi Z\alpha}{3}|\Psi(0)|^2\left[\langle r^2\rangle - \frac{Z\alpha m_r}{2}\langle r^3\rangle_{(2)} + (Z\alpha)^2 (F_\mathrm{REL} + m_r^2 F_\mathrm{NREL}) \right].
\end{equation}
\end{widetext}
Here, $\Psi(0)$ denotes the muon wave function at the origin,
\rr{} is the rms charge radius of the nucleus, and
$\langle r^3\rangle_{(2)}$ is its ``Friar moment''~\footnote{
$\langle r^3\rangle_{(2)}$ has been coined
``third Zemach moment'' by Friar~\cite{Friar:1978:Annals}.
To avoid confusion with the ``Zemach radius'' $r_Z$ that appears in the 
finite size effect in the 2S hyperfine splitting (Sec.~\ref{sec:HFS:Rz})
we adopt the term ``Friar moment'' as recently suggested by 
Karshenboim~\cite{Karshenboim:2015:PRD91_073003}.}.
As detailed in Sec.~\ref{sec:LS:Pol} there is no contribution from the 
Friar moment due to a cancellation with part of the inelastic deuteron
``polarizability'' contributions.\\
In Eq.~(\ref{eq:LS:FinSize}), 
$F_\mathrm{REL}$ and $F_\mathrm{NREL}$ contain various moments of the
nuclear charge distribution, see Ref.~\cite{Friar:1978:Annals}, and in
particular the Appendix~E therein for analytic expressions for some
simple model charge distributions.

\begin{figure}[h]
  \begin{center}
    \begin{fmffile}{ope}
      \begin{fmfgraph*}(80,80)
        \fmftop{i1,o1}
        \fmfbottom{i2,o2}
        \fmf{plain,tension=1.0}{i1,v1,o1}
        \fmf{plain,width=3}{i2,v2,o2}
        \fmf{photon,tension=0}{v1,v2}
        \fmfv{decor.shape=circle,decor.filled=full,decor.size=10}{v2}
        \fmfv{label=d}{i2}
        \fmfv{label.angle=-150,label=$\mu$}{i1}
      \end{fmfgraph*}
    \end{fmffile}
  \end{center}
  \caption{
      Item (r1), the leading nuclear finite size correction
      stems from a one-photon interaction with a deuteron form factor insertion,
      indicated by the thick dot.}
  \label{fig:OPE}
\end{figure}
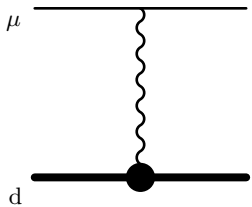

The leading order finite size effect, item (r1) in Tab.~\ref{tab:LS:Radius},
is the first term in Eq.~(\ref{eq:LS:FinSize}).
It originates from the one-photon exchange with a deuteron form factor 
insertion shown in Fig.~\ref{fig:OPE}
and is proportional to the rms charge radius of the deuteron, \rrd.

For (r2), the radiative correction $\alpha(Z\alpha)^5$, we chose Martynenko's value: 
The equations for the calculation of this term are given in \cite{Eides:1997:two-photon}.
Borie~\cite{Borie:2014:arxiv_v7}, Tab.\,14 uses
Eq.\,(10) of Ref.~\cite{Eides:1997:two-photon} which gives the total radiative correction of
order $\alpha(Z\alpha)^5$, i.e.\ the sum of Eqs.~(7) and (9) in Ref.~\cite{Eides:1997:two-photon}.
Martynenko~\cite{Krutov:2011:PRA84_052514}, in contrast, uses Eq.~(9) of 
Ref.~\cite{Eides:1997:two-photon}, stating that the additional polarization correction, 
Eq.~(7) in Ref.~\cite{Eides:1997:two-photon},
which is included in Eq.~(10), cancels with a part of the (inelastic) deuteron polarizability contribution.

\begin{table}[b]
  \renewcommand{\baselinestretch}{1.1}
  \renewcommand{\arraystretch}{1.5}
  \caption{The item (r3'), 
    {\em remaining $(Z\alpha)^6$ corrections to finite size}
    from Tab.~\ref{eq:LS:Radius}, in meV, evaluated
    for an exponential and a uniform charge distribution of the deuteron,
    using the formulas given by Borie~\cite{Borie:2014:arxiv_v7} p.~30,
    and Martynenko~\cite{Martynenko:2014:muD_Theory} Eq.~(33), 
    and the moments from Friar~\cite{Friar:1978:Annals} Tab.~V.
    For details, see text.
    Our average, $\Delta E\mathrm{(r3')} ~ = ~ 0.0030(6)$\,meV,
    is obtained from the spread of these values.}
  \label{tab:r3prime}
  \setlength\tabcolsep{1mm}
  \setlength{\extrarowheight}{0.2mm}
  \begin{center}
    \begin{tabular}{ l | c c}
      \hline
      \hline
      Distribution & after Borie~\cite{Borie:2014:arxiv_v7}  & after Martynenko~\cite{Martynenko:2014:muD_Theory} \\
      \hline
      Exponential  & 0.00238  & 0.00285   \\
      Uniform      & 0.00350  & 0.00355   \\
      \hline
      \hline
    \end{tabular}
  \end{center}
\end{table}
The finite size correction to the Lamb shift of order $(Z\alpha)^6$ has first
been calculated by Friar~\cite{Friar:1978:Annals}, see in particular Appendix~E
therein.
Both Borie~\cite{Borie:2014:arxiv_v7} (p.\,30) and 
Martynenko~\cite{Martynenko:2014:muD_Theory} (Eq.~(33)) 
follow Friar~\cite{Friar:1978:Annals} and evaluate this contribution 
as the sum of two terms which we list separately:\\
The first one, (r3), has an explicit $\langle r_d^2\rangle$ dependence, while
the second one, (r3'), is usually evaluated for an exponential charge 
distribution,
since a model-independent evaluation of this term is prohibitively 
difficult~\cite{Borie:2014:arxiv_v7}.
Small differences between the formulas given by 
Borie~\cite{Borie:2014:arxiv_v7} and 
Martynenko~\cite{Martynenko:2014:muD_Theory} result in values for (r3') of
0.0029\,meV and 0.0031\,meV.
For example, the term {\em $\langle r^2 \rangle \langle ln(\mu r) \rangle$},
which is part of $\mathrm{F_{REL}}$
in Eq.~(\ref{eq:LS:FinSize})~\footnote{See e.g.\ Ref.~\cite{Friar:1978:Annals} Eq.~(43) or Ref.~\cite{Borie:2014:arxiv_v7} p.~30.}, 
is attributed to (r3) and (r3') by Martynenko and Borie, respectively.

We calculate (r3') from Borie's and Martynenko's formula, 
for both an exponential and a uniform charge distribution,
using the moments given by Friar~\cite{Friar:1978:Annals}
and obtain the values listed in Tab.~\ref{tab:r3prime}.
We adopt the average, $\Delta E\mathrm{(r3')} ~ = ~ 0.0030(6)$\,meV.

The items (r4) and (r5) do not depend on the shape of the deuteron charge 
distribution~\cite{Borie:2014:arxiv_v7,Karshenboim:PC:2015}.
The two-loop vacuum polarization corrections (r6) and (r7) are 
only given by Martynenko~\cite{Martynenko:2014:muD_Theory}.

A correction to the 2P$_{1/2}$
level (r8) is given by Borie~\cite{Borie:2014:arxiv_v7}.
Item (r8) shifts the $2P_{1/2}$ level ``upwards'' (less bound). This
{\em increases} the energy difference between the $2S$ and $2P_{1/2}$ levels,
which explains the positive sign of this contribution in
Tab.~\ref{tab:LS:Radius}.
At the same time, this term {\em decreases} the fine structure
($2P_{3/2} - 2P_{1/2}$ energy difference) and is hence listed as item (f10)
with a negative sign in Tab.~\ref{tab:fs}.

The total radius-dependent contribution to the Lamb shift yields
\begin{equation}
\begin{aligned}
  \label{eq:LS:Radius}
  \Delta E^\mathrm{LS}_\mathrm{rad.-dep.} = &
  -6.11025(28) ~ \rd^2 ~~ \mathrm{meV/fm^2} \\
  & + 0.00300(60)\,\mathrm{meV}.
\end{aligned}
\end{equation}

%
\onecolumngrid
\onecolumngrid

\begin{table}[t!]
  \onecolumngrid
  \footnotesize
  \setlength\extrarowheight{3pt}
  \centering
  \caption{Coefficients of the {\bf radius-dependent} contributions to the Lamb shift.
    Values are in meV/fm$^2$, except for {r3'}~$^{\rm e}$.
    KS: K\"all\'en-Sabry, VP: vacuum polarization, SOPT: second-order perturbation theory.
  }
  \label{tab:LS:Radius}
  \hspace*{-10mm}\begin{tabular}{l|l |f{6}    c     |f{5}   c    |f{4}      c   | f{7}@{ }f{5} c}
    \hline
    \hline
       & Contribution  & \cntl{2}{Martynenko}
                                     & \cntl{2}{Borie}
                                                   & \cntl{2}{Karshenboim}
                                                                  & \cnt {3}{Our choice} \\
       &               & \cntl{2}{\cite{Martynenko:2014:muD_Theory}} 
                                     & \cntl{2}{\cite{Borie:2014:arxiv_v7} Tab.14} 
                                                   & \cntl{2}{\cite{Karshenboim:2012:PRA85_032509} Tab.III}
                                                                  & \cnt{2}{value}
                                                                                         &  \cnt{1}{source}   \\
    \hline
    r1 & Leading fin.\ size corr., $(Z\alpha)^4$
                       & -6.07313                                         & (27)  
                                     & -6.0730                            & $b_a$  
                                                   & -6.0732              & $\Delta E_{FNS}^{(0)}$
                                                                  &  -6.07310 
                                                                          & \pm 0.00010
                                                                              & avg. \\
    r2 & Radiative corr., $\alpha(Z\alpha)^5$
                       & -0.000962~\footnote{This value was published with a wrong sign in~\cite{Krutov:2011:PRA84_052514}.
      The term is from Eq.~(9) in \cite{Eides:1997:two-photon}.}
                                                                          & (62), \cite{Krutov:2011:PRA84_052514} 
                                     & -0.00072~\footnote{This value is obtained in \cite{Eides:1997:two-photon}, Eq.~(10). 
                                                  For further explanations see Sec.\,\ref{sec:LS:Radius}.}
                                                                          & $b_b$  
                                                   &                      & 
                                                                  &  -0.000962&           
                                                                             & M \\
    r3 & Finite size corr. order $(Z\alpha)^6$
                       & -0.002128                                        & (33)  
                                     & -0.00212                           & $b_c$  
                                                   &                      & 
                                                                  &  -0.002124& \pm 0.000004
                                                                             & avg. \\
    r4 & Uehling corr.\ (+KS), $\alpha(Z\alpha)^4$
                       & -0.01350                                         & (28)  
                                     & -0.0130                            & $b_d$  
                                                   & -0.0132              & $\Delta E_{FNS}^{(2)}$
                                                                  &  -0.01325& \pm 0.00025
                                                                             & avg. \\
    r5 & One-loop VP in SOPT, $\alpha(Z\alpha)^4$
                       & -0.020487                                        & (29)  
                                     & -0.02062                           & $b_e$  
                                                   & -0.0205              & $\Delta E_{FNS}^{(1)}$
                                                                  & -0.020554& \pm 0.000067
                                                                             & avg. \\
    r6 &Two-loop VP corr., $\alpha^2(Z\alpha)^4$
                       & -0.000105                                        & (30, 31)  
                                     &                                    &        
                                                   &                      & 
                                                                  & -0.000105 &           
                                                                             & M \\
    r7 & Two-loop VP in SOPT, $\alpha^2(Z\alpha)^4$
                       & -0.000095                                        & (32)  
                                     &                                    &        
                                                   &                      & 
                                                                  & -0.000095 &           
                                                                             & M \\
    r8 & Corr.\ to the $2P_{1/2}$ level
                       &                                                  &       
                                     & -0.0000606                         & $b(2p_{1/2})$  
                                                   &                      & 
                                                                  & +0.0000606~\footnote{The sign is explained in the text.}&
                                                                             & B\\
    \hline
       & Sum           & -6.10848                                         &       
                                     & -6.10952~\footnote{The \rr{} coefficient given in Ref.~\cite{Borie:2014:arxiv_v7} page 13, neglects the {\em correction to the {\rm 2P$_{1/2}$} level}, item (r8).} & 
                                                   & -6.1069              & $\Delta E_{FNS}$
                                                                  & -6.11013 & \pm 0.00028
                                                                             &      \\
    \hline
    \lft{11}{~} \\
    \hline
    r3'& Remaining order $(Z\alpha)^6$\,[meV]~\footnote{Belongs to r3. Depends on the charge distribution in a non-trivial way, see text.}
                       &  0.0029\mathrm{~meV}                             & (33)  
                                     &  0.0033\mathrm{~meV}               &        
                                                   &                      & 
                                                                  & 0.00300   & \pm 0.00060\mathrm{~meV}
                                                                             & Tab.~\ref{tab:r3prime} \\
    \hline
    \hline
    &&&&&&&&&& \\
       & \bf Sum
                       & \multicolumn{2}{l|}{-6.10848~$\rd^2$ + 0.0029\,meV}                
                                     & \multicolumn{2}{l|}{-6.10952~$\rd^2${} + 0.0033\,meV}  
                                                   &                      &
                                                                  & \multicolumn{3}{l}{\bf -6.11025(28)~$\boldsymbol \rd^2${} + 0.00300(60)\,meV} \\
    &&&&&&&&&& \\
    \hline
    \hline
  \end{tabular}
\end{table}
\twocolumngrid

\mbox{~}
\clearpage

\subsection{Nuclear polarizability contributions to the Lamb shift}
\label{sec:LS:Pol}

\begin{figure}[b]
  \setlength{\unitlength}{0.0040\columnwidth}
  \begin{center}
    \begin{minipage}{0.45\columnwidth}
      (a)\hfill\mbox{~}\\
      \vspace{0.5ex}
      \begin{fmffile}{tpe_elastic_1}
        \begin{fmfgraph*}(80,60)
          \fmftop{i1,o1}
          \fmfbottom{i2,o2}
          \fmf{plain}{i1,t1,txx,t2,o1}
          \fmf{plain,width=3}{i2,b1,bb,b2,o2}
          \fmfv{decor.shape=circle,decor.filled=full,decor.size=10}{b1}
          \fmfv{decor.shape=circle,decor.filled=full,decor.size=10}{b2}
          \fmf{photon,tension=0}{t1,b1}
          \fmf{photon,tension=0}{b2,t2}
          \fmfv{label=d}{i2}
          \fmfv{label.angle=-150,label=$\mu$}{i1}
        \end{fmfgraph*}
      \end{fmffile}
    \end{minipage}
    \hfill
    \begin{minipage}{0.45\columnwidth}
      (c)\hfill\mbox{~}\\
      \vspace{0.5ex}
      \begin{fmffile}{tpe_inelastic_1}
        \begin{fmfgraph*}(80,60)
          \fmftop{i1,o1}
          \fmfbottom{i2,o2}
          \fmf{plain}{i1,t1,txx,t2,o1}
          \fmf{plain,width=3}{i2,b1}
          \fmf{plain,width=3}{b2,o2}
          \fmfpoly{smooth,filled=30, pull=1.4, tension=0.2}{b1,b10,b2,b11}
          \fmffreeze
          \fmfshift{14up}{b10}
          \fmfshift{14down}{b11}
          \fmf{photon,tension=0}{t1,b1}
          \fmf{photon,tension=0}{b2,t2}
          \fmfv{label=d}{i2}
          \fmfv{label.angle=-150,label=$\mu$}{i1}
        \end{fmfgraph*}
      \end{fmffile}
    \end{minipage}
    \vspace{6ex}

    \begin{minipage}{0.45\columnwidth}
      (b)\hfill\mbox{~}\\
      \vspace{0.5ex}
      \begin{fmffile}{tpe_elastic_2}
        \begin{fmfgraph*}(80,60)
          \fmftop{i1,o1}
          \fmfbottom{i2,o2}
          \fmf{plain}{i1,t1,txx,t2,o1}
          \fmf{plain,width=3}{i2,b1,bb,b2,o2}
          \fmfv{decor.shape=circle,decor.filled=full,decor.size=10}{b1}
          \fmfv{decor.shape=circle,decor.filled=full,decor.size=10}{b2}
          \fmf{photon,tension=0}{t1,b2}
          \fmf{photon,tension=0}{b1,t2}
          \fmfv{label=d}{i2}
          \fmfv{label.angle=-150,label=$\mu$}{i1}
        \end{fmfgraph*}
      \end{fmffile}
    \end{minipage}
    \hfill
    \begin{minipage}{0.45\columnwidth}
      (d)\hfill\mbox{~}\\
      \vspace{0.5ex}
      \begin{fmffile}{tpe_inelastic_2}
        \begin{fmfgraph*}(80,60)
          \fmftop{i1,o1}
          \fmfbottom{i2,o2}
          \fmf{plain}{i1,t1,txx,t2,o1}
          \fmf{plain,width=3}{i2,b1}
          \fmf{plain,width=3}{b2,o2}
          \fmfpoly{smooth,filled=30, pull=1.4, tension=0.2}{b1,b10,b2,b11}
          \fmffreeze
          \fmfshift{14up}{b10}
          \fmfshift{14down}{b11}
          \fmf{photon,tension=0}{t1,b2}
          \fmf{photon,tension=0}{b1,t2}
          \fmfv{label=d}{i2}
          \fmfv{label.angle=-150,label=$\mu$}{i1}
        \end{fmfgraph*}
      \end{fmffile}
    \end{minipage}
  \end{center}
  \caption{\label{fig:tpe}
    (a)+(b) Elastic \ELSFriar{},  and (c)+(d) inelastic \ELSinelast{}
    two-photon exchange (TPE) contribution.
    The thick dots in (a) indicate deuteron form factor insertions.
    The blob in (c) and (d) represents all possible excitations of the nucleus.
    The elastic part (a)+(b) is canceled by a part of the inelastic
    (polarizability) contribution of (c)+(d)~\cite{Pachucki:2011:PRL106_193007,Hernandez:2014:PLB736_344},
    just like for electronic deuterium~\cite{Friar:1997:PRA56_5173,Friar:2013:PRC88_034004}.
  }
\end{figure}
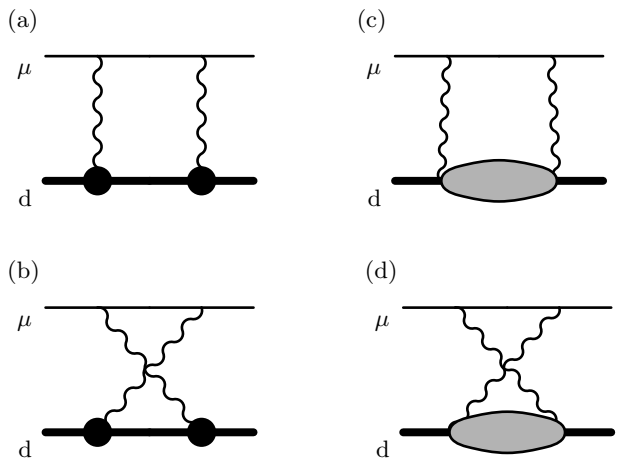

Historically, the two-photon exchange (TPE) contribution to the Lamb shift (LS)
in muonic atoms
has been considered the sum of the two parts displayed in Fig.~\ref{fig:tpe}
(a,b) and (c,d), respectively:
\begin{equation}
  \ELSTPE{} ~ = ~ \ELSFriar{} ~ + \ELSinelast{}
\end{equation}
The elastic ``Friar moment'' contribution, \ELSFriar{},
also known as ``third Zemach moment'' $\langle r^3\rangle_{(2)}$ contribution,
shown in Fig.~\ref{fig:tpe}(a,b) is sensitive to the shape of the nuclear charge
distribution, beyond the leading \rr{} dependence 
discussed in Sec.~\ref{sec:LS:Radius}. This part is traditionally
parameterized as being proportional to the third power of the rms charge
radius. 
The coefficient depends
on the assumed radial charge distribution. 
For example, Borie
gives $\langle r^3\rangle_{(2)} = 4.0(2)\,r^3$,
where $r=\surd\overline{\langle r^2 \rangle}$.
For \mud{}, the Friar (3rd Zemach) moment contribution
amounts to $\sim 0.43$\,meV~\cite{Borie:2014:arxiv_v7}.

The inelastic part, \ELSinelast{}, 
frequently termed ``nuclear polarizability contribution''
is shown in Fig.~\ref{fig:tpe}(c,d).  It stems from virtual 
excitations of the nucleus due to the exchange of two photons with the muon.
The inelastic contributions
are notoriously the least well-known theory contributions and limit the
extraction of the charge radius from laser spectroscopy of the Lamb shift.

Early calculations of the contribution from the deuteron polarizability, i.e.\
the inelastic part \ELSinelast{} displayed in Fig.~\ref{fig:tpe}(c,d) include
Fukushima \etal{}~\cite{Fukushima:1992:PRA46}, $1.24$\,meV,
Lu and Rosenfelder~\cite{Lu:1993:PLB319_with_erratum}, $1.45 \pm 0.06$\,meV,
and Leidemann and Rosenfelder~\cite{Leidemann:1995:PRC51_427}, $1.500 \pm 0.025$\,meV.
The latter value 
has been used extensively in the literature.

\subsubsection{Modern determinations of \ELSTPE{}}
Recently, several works have revisited the TPE contributions 
to the Lamb shift in \mud. 
Tab.~\ref{tab:LS:Pol} lists the contributions to
 \ELSTPE{} obtained by
Pachucki (2011)~\cite{Pachucki:2011:PRL106_193007},
Friar (2013)~\cite{Friar:2013:PRC88_034004},
Carlson \etal{} (2014)~\cite{Carlson:2014:PRA89_022504},
the TRIUMF/Hebrew University group in Hernandez \etal{} (2014)~\cite{Hernandez:2014:PLB736_344},
and Pachucki and Wienczek (2015)~\cite{Pachucki:2015:PRA91_040503}.

As it will turn out that the uncertainty in \ELSTPE{} is by far
the largest uncertainty in the determination of \rd{} from
the \mud{} data, we next summarize the main features of these papers.
We identify missing and incorrect terms in the original papers.
The detailed compilation in Tab.~\ref{tab:LS:Pol} 
allows us to obtain the reliable average
given in Eq.~(\ref{eq:LS:pol}).

In his 2011 paper~\cite{Pachucki:2011:PRL106_193007}, 
Pachucki calculated the nuclear structure corrections to the Lamb
shift in muonic deuterium using the
AV18 potential for the deuteron and obtained
\ELSTPE{} = 1.680(16)\,meV.
Moreover, he confirmed that for \mud{}, 
similar to electronic deuterium~\cite{Friar:1997:PRA56_5173,Friar:2013:PRC88_034004},
the elastic ``Friar moment'' contribution 
of order $(Z\alpha)^5$, \ELSFriar{} (Fig.~\ref{fig:tpe}(a,b))
is canceled by a part of the
inelastic two-photon (polarizability) contributions, \ELSinelast{}
(Fig.~\ref{fig:tpe}(c,d)).
The reason for this cancellation is that the deuteron binding energy of 
2.2\,MeV is small compared to the muon mass~\footnote{
For muonic hydrogen, in contrast, the first excited state of the nucleus (proton)
is the $\Delta$ resonance with an excitation energy of 300\,MeV. Hence
there is no such cancellation between elastic and inelastic TPE contributions
in \mup.}.\\
Pachucki~\cite{Pachucki:2011:PRL106_193007}
includes both the elastic and inelastic TPE contribution of the
proton, but not the neutron.
For the proton, he rescaled the full proton TPE contribution
calculated for muonic {\em hydrogen}~\cite{Carlson:2011:PRA84_020102},
$\Delta E(2S) = -0.0369(24)$\,meV,
with a reduced mass ratio to correct for the larger
wave function overlap in \mud{},
\begin{equation}
  \label{eq:wf_overlap}
  \zeta = (\,m_{\rm r}^{\mud}\,/\,m_{\rm r}^{\mup}\,)\,^3 ~ = ~ 1.1685.
\end{equation}
This gives a value of 0.043(3)\,meV for our items p13+p14.\\
Pachucki's value for the magnetic contribution (p10) was found to be wrong
by a factor of two in Ref.~\cite{Hernandez:2014:PLB736_344} and
was corrected in Pachucki's later work~\cite{Pachucki:2015:PRA91_040503}.

Friar (2013) 
used the zero-range approximation (ZRA)~\cite{Friar:2013:PRC88_034004}
which allows for a systematic derivation of all terms.
Friar finds very good agreement with the
results of Pachucki~\cite{Pachucki:2011:PRL106_193007} despite the simplicity of
the ZRA. The cancellation between elastic and inelastic TPE contributions
is observed in ZRA, too~\footnote{Friar's paper~\cite{Friar:2013:PRC88_034004}
gives a very good understanding about the nature of this cancellation.}.
Friar noted that a nucleon finite size contribution of 0.029\,meV
should be added that had not been included in Ref.~\cite{Pachucki:2011:PRL106_193007}.
Friar's value of \ELSTPE{} = 1.941 $\pm$ 1\% \,meV seems at first glance to be in
serious disagreement with Pachucki's value~\cite{Pachucki:2011:PRL106_193007}.
The difference is however mainly caused by the Coulomb distortion (p5+p6) of $-0.263$\,meV
which should be included in every calculation~\cite{Gorchtein:PC:2015}.
Including further items in Tab.~\ref{tab:LS:Pol} like 
the nucleon polarizability contribution p14+p15, and 
the nucleon subtraction term p16
results in a 
``corrected value'' of 1.697\,meV.
Higher order corrections to the dipole contribution (p3 and p4)
can account for the remaining difference to the other model calculations
and ``our avg.''.

In their 2014 paper~\cite{Hernandez:2014:PLB736_344} the 
TRIUMF/Hebrew University group
performed an independent calculation
using two parameterizations of the deuteron potential:
AV18 one the one hand, 
and nucleon-nucleon (NN) forces from chiral effective field theory ($\chi$EFT) 
up to order N$^3$LO and with various cutoffs, on the other.
As in their work on muonic helium-4~\cite{Ji:2013:PRL111,Ji:2014:FewBodySyst}
they added higher order relativistic corrections,
corrected the magnetic term,
and added the intrinsic neutron polarizability~\cite{Bacca:PC:2015}.
They also introduced the reduced-mass-dependence in higher-order
terms, while the earlier Ref.~\cite{Pachucki:2011:PRL106_193007} 
had worked in the limit of infinite nuclear mass for all terms.
The nuclear mass dependence for all terms was then further refined 
in Ref.~\cite{Pachucki:2015:PRA91_040503}.\\
Ref.~\cite{Hernandez:2014:PLB736_344} observed the cancellation of 
elastic and inelastic contribution for the muonic deuterium case explicitly:
The sum of terms $\delta^{(1)}_{Z1} + \delta^{(1)}_{Z3} = -0.424(3)$\,meV~\footnote{The
sign convention in Ref.~\cite{Hernandez:2014:PLB736_344}
is opposite to the one used here.}
in Ref.~\cite{Hernandez:2014:PLB736_344}
cancels very nicely with elastic Friar (``3rd Zemach'') contribution
\ELSFriar{} = 0.433(21)\,meV of the deuteron
as calculated by Borie, see Ref.~\cite{Borie:2014:arxiv_v7} p.~7.\\
Averaging over their results 
from AV18 and N$^3$LO they obtained a value of 
\ELSTPE{} = 1.690(20)\,meV.
The apparently good agreement with Pachucki's
value~\cite{Pachucki:2011:PRL106_193007}
may however be accidental as it arises 
from the cancellation of
many small differences~\cite{Hernandez:2014:PLB736_344,Bacca:PC:2015}.
Again, adding omitted items (p13 and p16) results
in very good agreement with all other sums in Tab.~\ref{tab:LS:Pol}.
Note that the 4th column in our Tab.~\ref{tab:LS:Pol} (``Source 4'') 
is that
value from Tab.~3 of Ref.~\cite{Hernandez:2014:PLB736_344}, 
columns ``N$^3$LO-EM'' and ``N$^3$LO-EGM'', which deviates most
from their AV18 result.
This is an attempt to be rather conservative when 
determining ``our average'' following Eq.~(\ref{eq:avg}).\\
According to Bacca~\cite{Bacca:PC:2015},
their values for $\delta^{(2)}_{NS}$ (our item p11) should be updated
to $+0.020$\,meV from the published value of $+0.015$\,meV
\cite{Hernandez:2014:PLB736_344}.

The 2015 paper by Pachucki and Wienczek~\cite{Pachucki:2015:PRA91_040503}
updated Pachucki's results from 2011 (Ref.~\cite{Pachucki:2011:PRL106_193007}),
again using the AV18 potential.
Among other things, they included
the finite size of the nucleons, and 
the intrinsic elastic and inelastic two-photon exchange with individual
nucleons.
They corrected their magnetic interaction term,
and derived the correct mass dependence of the TPE correction
and its consistent separation with the so-called pure recoil correction.
Their total TPE contribution of order $(Z\alpha)^5$ is
$1.717 \pm 0.020$\,meV.
Item p16 must be added to obtain a ``corrected value''
in very good agreement with all other determinations.

Complementary to the calculations using various deuteron 
potentials~\cite{Pachucki:2011:PRL106_193007,Friar:2013:PRC88_034004,Hernandez:2014:PLB736_344,Pachucki:2015:PRA91_040503},
Carlson \etal{}~\cite{Carlson:2014:PRA89_022504}, in 2014, determined
the TPE contributions with minimal model dependence 
using measured elastic and inelastic electron-deuteron scattering data and
dispersion relations.
Their model-independent calculation yields 2.01(74)\,meV, confirming the numbers
given by \cite{Pachucki:2011:PRL106_193007,Friar:2013:PRC88_034004,Hernandez:2014:PLB736_344}
albeit with a much larger uncertainty.
This uncertainty stems from the uncertainty in the data
and can be improved significantly when new data 
from the Mainz MAMI and MESA facilities
becomes available~\cite{Carlson:2014:PRA89_022504}.\\
The several contributions to their sum can not be easily equated
with individual items p1...p16 listed in Tab.~\ref{tab:LS:Pol}, so we do not
quote their individual contributions, with one important exception.
Carlson \etal{}~\cite{Carlson:2014:PRA89_022504}
note that the proton and neutron intrinsic polarizabilities 
of 
0.028(2)\,meV (our items p14+p15) should be added to the 
earlier results of 
Pachucki~\cite{Pachucki:2011:PRL106_193007} and Friar~\cite{Friar:2013:PRC88_034004}.
Such a correction is already included in
the later paper by Hernandez \etal{} \cite{Hernandez:2014:PLB736_344}.\\
On the other hand, the value of
Carlson \etal{}~\cite{Carlson:2014:PRA89_022504}
should be corrected~\cite{Gorchtein:PC:2015,BirseMcGovern:PC:2015}
for Coulomb distortion (p5+p6) of $-0.263$\,meV.
Then the central value becomes 1.748\,meV, in even better agreement
with the (corrected) values from nuclear 
models~\cite{Pachucki:2011:PRL106_193007,Friar:2013:PRC88_034004,Hernandez:2014:PLB736_344,Pachucki:2015:PRA91_040503}.
Note that this correction corresponds to 1/3 of their quoted uncertainty
and may hence look absurd. But the uncertainty quoted in 
Ref.~\cite{Carlson:2014:PRA89_022504}
originates almost exclusively from the plain wave Born approx.\ (PWBA)
term and may be reduced by at least a factor of 4 with
new data from a planned experiment in Mainz~\cite{Carlson:2014:PRA89_022504}.
The good agreement between the corrected central value
and all other (corrected) values makes one wonder if the
uncertainty in Ref.~\cite{Carlson:2014:PRA89_022504} is
maybe somewhat conservative.\\
The ``Thomson term'' is a recoil correction that has first been
calculated in Ref.~\cite{Carlson:2014:PRA89_022504}. It has received some
attention~\cite{Gorchtein:PC:2015,BirseMcGovern:PC:2015,Pachucki:PC:2015,Bacca:PC:2015}
in the discussion of our Tab.~\ref{tab:LS:Pol}, and the conclusion was that 
this term is indeed correctly added to the sum of the contributions in the 
dispersion-relation treatment of Ref.~\cite{Carlson:2014:PRA89_022504}. 
The other calculations~\cite{Pachucki:2011:PRL106_193007,Friar:2013:PRC88_034004,Hernandez:2014:PLB736_344,Pachucki:2015:PRA91_040503}
have correctly {\em not} included such a term,
because the cancellation between elastic and (part of the) inelastic 
contributions to the polarizability will eliminate this Thomson term 
(as well as other similar recoil-like terms)
in such a ``nuclear Hamiltonian approach''.
All further recoil corrections of order $(Z \alpha)^5$ 
to the Lamb shift in muonic deuterium
are then included in the ``pure recoil corrections'', item \#22 in 
Tab.~\ref{tab:LS:QED}.

\subsubsection{Comparison of terms and further corrections}

An earlier version of the present manuscript 
was sent to the authors of 
Refs.~\cite{Pachucki:2011:PRL106_193007,Friar:2013:PRC88_034004,Carlson:2014:PRA89_022504,Hernandez:2014:PLB736_344,Pachucki:2015:PRA91_040503}
and other experts in the field. The ensuing insightful discussions
resolved several discrepancies between the published values of \ELSTPE{}
and revealed that some further corrections
should be included.

Table~\ref{tab:LS:Pol} lists in chronological order 
the modern determinations of \ELSTPE{} using 
various nuclear models, and scattering data.
As usual, we calculate an ``average'' following our Eq.~(\ref{eq:avg})
and consider the spread of values in the uncertainty.

Items {\bf p1 through p10} contain the nuclear contributions, and 
the various calculations are in good agreement.

It is satisfying to note that the dominant dipole term, item p1,
is in very good agreement for the three models used: 
AV18, ZRA, and N$^3$LO $\chi$EFT.
We average the results using ``modern'' 
potentials~\cite{Hernandez:2014:PLB736_344,Pachucki:2015:PRA91_040503} and
take the agreement of the ZRA result as an indication that the ZRA results for
the smaller terms are likely to be accurate on the few $\mu$eV level and
can hence be used in ``our average''.

Items p2..p4 are relativistic corrections to p1. The two most recent
works~\cite{Hernandez:2014:PLB736_344,Pachucki:2015:PRA91_040503}
include higher order relativistic corrections so we consider only
these works in the average.

There is consensus that the Coulomb distortion contribution (p5+p6)
should be included in {\em all} calculations.
Adding our average of -0.263\,meV to the results of 
Friar~\cite{Friar:2013:PRC88_034004} and 
Carlson \etal{}~\cite{Carlson:2014:PRA89_022504}
removes most of the discrepancy between all published values.

Nuclear excitation corrections p7..p9 cancel to some degree. Our average 
includes the results from ZRA~\cite{Friar:2013:PRC88_034004} and the 
most recent AV18 and N$^3$LO models~\cite{Hernandez:2014:PLB736_344,Pachucki:2015:PRA91_040503}.

The magnetic contribution p10 from Ref.~\cite{Pachucki:2011:PRL106_193007}
has been corrected in the later work~\cite{Pachucki:2015:PRA91_040503}.
We average over the other results.

Items {\bf p11 through p16} are the nucleon contributions.

Item p11 from the TRIUMF/Hebrew University group ($\delta^{(2)}_{NS}$
in Ref.~\cite{Hernandez:2014:PLB736_344}) has been updated to
to $+0.020$\,meV~\cite{Bacca:PC:2015},
further improving the
agreement with Refs.~\cite{Friar:2013:PRC88_034004,Pachucki:2015:PRA91_040503}.

After some discussions,
consensus has been reached that several nucleon contributions
should be included~\cite{Bacca:PC:2015,Pachucki:PC:2015,Borie:PC:2015,BirseMcGovern:PC:2015,Gorchtein:PC:2015}:
The elastic Friar (3rd Zemach) term of the proton (p13),
the inelastic proton and neutron contributions (p14+p15),
and the subtraction terms from both the proton and the neutron (p16)
are therefore included in our sum.
In principle, the elastic Friar term of the neutron should be included too,
but it is small enough to be 
neglected~\cite{Friar:2013:PRC88_034004,Pachucki:2015:PRA91_040503,BirseMcGovern:PC:2015}.

We follow the suggestion of Birse and McGovern~\cite{BirseMcGovern:PC:2015} 
who obtain these values as follows:

Item p13, the elastic Friar (3rd Zemach) moment contribution of the {\em proton}
to the Lamb shift in \mud{}
is obtained from the values of the elastic and the non-pole Born term calculated
for muonic hydrogen (\mup{})~\cite{BirseMcGovern:PC:2015}.\\
Both,
the elastic term in \mup{}, and the
non-pole term in \mup{},
have been obtained by Carlson and Vanderhaeghen 
from scattering data using dispersion relations~\cite{Carlson:2011:PRA84_020102}.
Their value for the elastic term, $\Delta E^{\rm el}$, 
amounts to  $0.0295 \pm 0.0013$\,meV~\footnote{The
sign convention in Ref.~\cite{Carlson:2011:PRA84_020102}
is opposite to the one used here.}.
Their value for the non-pole Born term for \mup{}
is
$-0.0048$\,meV~\footnotemark[9]. 
The sum of these two terms, rescaled with $\zeta$ from Eq.~(\ref{eq:wf_overlap})
yields for p13, the elastic Friar moment contribution of the proton to 
the Lamb shift in muonic deuterium, the value
\begin{equation}
  \label{eq:pol:protonFriar}
  \ELSFriar({\rm p}) = 0.0289 \pm 0.0015\,\mathrm{meV}.
\end{equation}
  
The inelastic proton and neutron polarizabilities p14 and p15
have been calculated from deuteron data and dispersion relations by 
Carlson \etal~\cite{Carlson:2014:PRA89_022504}.
Their result for the sum p14+p15 amounts to 
\begin{equation}
  \label{eq:pol:pn_inelastic}
  \ELSinelast({\rm p}) +
  \ELSinelast({\rm n}) = \\ 
  0.028 \pm 0.002\,\mathrm{meV}
\end{equation}
which is the value we adopt.
Hernandez~\etal~\cite{Hernandez:2014:PLB736_344} used
the value 0.027(2)\,meV from the same Ref.~\cite{Carlson:2014:PRA89_022504}.
This number is, however, only an estimate using numbers rescaled from muonic
hydrogen, whereas our choice Eq.~(\ref{eq:pol:pn_inelastic}) is
calculated from deuteron data, and the value in Eq.~(\ref{eq:pol:pn_inelastic})
should be used~\cite{Gorchtein:PC:2015}.

Finally, the contribution from the ``subtraction term'' of the nucleon 
polarizabilities has to be considered, too~\cite{BirseMcGovern:PC:2015,Gorchtein:PC:2015}. 
Birse and McGovern have calculated
the subtraction term for the 
inelastic TPE of the proton in muonic {\em hydrogen},
$\Delta E_{\rm sub} = -0.0042 \pm 0.0010$\,meV~\footnote{The sign convention in
Ref.~\cite{BirseMcGovern:2012} is opposite to the one used here.},
using chiral perturbation theory~\cite{BirseMcGovern:2012}. 
This value is in good agreement with the value
$\Delta E^{\rm subt} = -0.0053 \pm 0.0019$\,meV~\footnotemark[9] 
from Carlson and Vanderhaeghen~\cite{Carlson:2011:PRA84_020102} 
which was however obtained from a particular model of the proton form 
factor and an older value of the proton magnetic
polarizability~\cite{BirseMcGovern:PC:2015}.
For the deuteron, we hence adopt the former value, 
double it assuming 
that the proton and neutron contributions are approximately
the same~\cite{BirseMcGovern:PC:2015},
and rescale with $\zeta$ from Eq.~(\ref{eq:wf_overlap}) to
yield p16 for muonic deuterium
\begin{equation}
  \label{eq:pol:pn_subtr}
  \Delta E^{\rm LS}_{\rm sub}({\rm p}) +
  \Delta E^{\rm LS}_{\rm sub}({\rm n}) =
  -0.0098 \pm 0.0098\,\mathrm{meV}.
\end{equation}
Here we have assigned a 100\% uncertainty.

\subsubsection{Our choice}

Summing all values in Tab.~\ref{tab:LS:Pol}, and adding the
uncertainties from (the spreads of) our averaging in quadrature, gives 
\begin{equation}
  \label{eq:LS:pol_naive}
  \ELSTPE{}\,{\rm (simple)} = \POLVALOUR{} \pm \POLERROUR~\mathrm{meV}.
\end{equation}
This uncertainty is smaller than the published uncertainties in
all original papers~\cite{Pachucki:2011:PRL106_193007,Friar:2013:PRC88_034004,Carlson:2014:PRA89_022504,Hernandez:2014:PLB736_344,Pachucki:2015:PRA91_040503}.
Hence we
increase conservatively the uncertainty in our average to the
0.020\,meV obtained by the two most recent model 
calculations~\cite{Hernandez:2014:PLB736_344,Pachucki:2015:PRA91_040503}.

The total TPE contribution
of order $(Z\alpha)^5$ to the Lamb shift in muonic
deuterium~\footnote{Note 
that non-perturbative Coulomb corrections of higher order in
$(Z\alpha)$ have been accounted here, whereas the pure recoil part of
the TPE has been separately given in Tab.~\ref{tab:LS:QED} \#22.}
is hence
\begin{equation}
  \label{eq:LS:pol}
  \ELSTPE{}\,{\rm (final)} = \POLVALRND{} \pm \POLERRRND{}~\mathrm{meV}.
\end{equation}
Further rounding is deferred to Eq.~(\ref{eq:LS:full}).
%

The uncertainty of the TPE contribution is by far the dominant one, 
and it limits severely the accuracy of
the deuteron rms charge radius obtained from laser spectroscopy of 
the Lamb shift in muonic deuterium.

\clearpage
\begin{turnpage}
\begin{table}[h]
  \renewcommand{\baselinestretch}{1.1}
  \renewcommand{\arraystretch}{1.0}
  \mbox{~}\vspace{-18ex}
  \setlength\extrarowheight{3pt}
  \caption{Deuteron structure contributionsto the Lamb shift in muonic deuterium.
    Values are in meV.
    For source 4, the N$^3$LO $^\dagger$ calculation by Hernandez \etal{}~\cite{Hernandez:2014:PLB736_344} we use
    their value from the rightmost two columns of their Tab.~3 that differs most from their ``AV18'' value.
    Items with a diamond $^\diamondsuit$ are corrected from the published values, see footnotes.}
  \label{tab:LS:Pol}
  \setlength\tabcolsep{1mm}
  \begin{center}
    \fontsize{7pt}{7pt}\selectfont
    \vspace{-2ex}
  \hspace*{-12mm}
\begin{tabular}{c | l    |f{6}    l     |f{4}   l    |f{5}  f{5} l  |f{6}      l  || f{5} l   || f{5}@{ }f{5} c}
    \hline
    \hline
      Item & Contribution  & \cntl{2}{Pachucki~\cite{Pachucki:2011:PRL106_193007}}
                                  & \cntl{2}{Friar~\cite{Friar:2013:PRC88_034004}}
                                             & \cntl{3}{Hernandez \etal~\cite{Hernandez:2014:PLB736_344}}
                                                        & \cntll{2}{Pach.\&\,Wienczek~\cite{Pachucki:2015:PRA91_040503}}
                                                                  & \cntll{2}{Carlson \etal~\cite{Carlson:2014:PRA89_022504}}
                                                                           & \cnt {3}{Our choice} \\
      &               & \cnt{1}{AV18} & \cntl{1}{~}
                                  & \cnt{1}{ZRA} & \cntl{1}{~}
                                             &\cnt{1}{AV18} & \cnt{1}{N$^3$LO $^\dagger$} & \cntl{1}{~\hspace{9mm}~} 
                                                        & \cnt{1}{AV18} & \cntll{1}{~}
                                                                   & \cnt{1}{\hspace{4mm}data\hspace{3mm}~} & \cntll{1}{~\hspace{9mm}~}
                                                                           & \lft{3}{~\hfill \mbox{~value} \hfill \mbox{source\hspace{-8mm}}}\\
    \hline
     &  Source          & \cnt{1}{1} &
                                 & \cnt{1}{2} & 
                                         & \cnt{1}{3} & \cnt{1}{~ 4} & 
                                                             & \cnt{1}{5} &
                                                                     & \cnt{1}{6} & &\\
\hline
    \hline
    \end{tabular}
\end{center}
  \end{table}
\end{turnpage}
\clearpage

\subsection{Total Lamb shift in muonic deuterium}
\label{sec:LS_total}

Collecting the
radius-independent (mostly) QED contributions 
listed in Tab.~\ref{tab:LS:QED} and 
summarized in Eq.~(\ref{eq:LS:QED}),
the radius-dependent contributions 
listed in Tab.~\ref{tab:LS:Radius} and 
summarized in Eq.~(\ref{eq:LS:Radius}),
and the complete two-photon (polarizability) contribution \ELSTPE{}
from Eq.~(\ref{eq:LS:pol}),
we obtain for the $\mathrm{2S-2P}$ energy difference in muonic deuterium
\begin{widetext}
\begin{equation}
  \label{eq:LS:full}
  \begin{aligned}
    \Delta E(2S-2P_{1/2}) ~ = & ~\, \LSVALERR\,\mathrm{meV} \\
    & + ~ 0.00300(60)\,\mathrm{meV}
    ~ - ~ 6.11025(28) ~ \rd^2 ~ \mathrm{meV/fm^2}\\
    & + ~ 1.70910(2000)\,\mathrm{meV}\\
    = &  ~ \, 230.486(20)~\mathrm{meV} - 6.1103(3)~ \rd^2 ~ \mathrm{meV/fm^2}
  \end{aligned}
\end{equation}
\end{widetext}
where in the last step we have rounded the values to reasonable accuracies.
%

One should note that the uncertainty of \POLERRFINAL\,meV 
from the nuclear structure corrections \ELSTPE{},
Eq.~(\ref{eq:LS:pol}), 
is about 30 times larger than the combined uncertainty of all 
radius-independent terms summarized in Tab.~\ref{tab:LS:QED},
and 15 times larger than the  uncertainty in the \rr{} coefficient 
(which amounts to 0.0013\,meV).
A further improvement of the nuclear structure corrections in 
light muonic atoms is therefore desirable.

\section{2S hyperfine splitting}
\label{sec:HFS}

\subsection{Fermi and Breit contributions}
The interaction between the magnetic moment of the nucleus with the
magnetic field induced by the lepton gives rise to shifts and
splittings of the energy levels termed hyperfine effects. In classical
electrodynamics, the interaction between the magnetic moments $\boldsymbol{\mu}_\mathrm{d}$ and
$\boldsymbol{\mu}_\mu$ of deuteron and muon, respectively, is described by \cite{Eides:2001:PhysRep}
\begin{equation}\label{eq:ham}
  H_\mathrm{HFS}^\mathrm{classical}=-\frac{2}{3}\boldsymbol{\mu}_\mathrm{d} \cdot\boldsymbol{\mu}_\mu \delta (\boldsymbol{r})
\end{equation}
where $\delta (\boldsymbol{r})$ is the delta-function in coordinate space.
A similar Hamiltonian to the one in Eq.\,(\ref{eq:ham}) can be derived in 
quantum field theory from the one-photon exchange diagram. Using the Coulomb 
wave function, this gives rise in first-order perturbation theory to an 
energy shift for muonic deuterium $n$S-states of~\cite{Borie:2014:arxiv_v7}
\begin{equation}
\begin{aligned}
\label{eq:HFS}
  E_\mathrm{HFS}(F) &=\frac{4(Z\alpha)^4m_r^3}{3n^3 m_\mu
    m_\mathrm{d}}(1+\kappa)(1+a_\mu)\frac{1}{2}\left[ F(F+1)-\frac{11}{4}\right] \\
                  &= \frac{1}{3} \Delta E_\mathrm{Fermi} \left[ F(F+1)-\frac{11}{4} \right]
\end{aligned}
\end{equation}
where $\Delta E_\mathrm{Fermi}$ is the Fermi splitting, $m_\mathrm{d}$ is the deuteron
mass, $F$ is the total angular momentum, $\kappa$ and $a_\mu$ are the deuteron
and muon anomalous magnetic moments, respectively.

The Fermi splitting
\begin{equation}
\begin{aligned}
\label{eq:EFermi}
  \Delta E_\mathrm{Fermi} &=  \frac{2(Z\alpha)^4m_r^3}{n^3 m_\mu
    m_\mathrm{d}}(1+\kappa)(1+a_\mu)\\
  &= \frac{3}{2} \beta_D (1+a_\mu)\\
\end{aligned}
\end{equation}
with
\begin{equation}
\begin{aligned}
\label{eq:betaD}
  \beta_D &= \frac{4(Z\alpha)^4m_r^3}{3n^3 m_\mu m_\mathrm{d}}(1+\kappa)
\end{aligned}
\end{equation}
is the main contribution to the HFS, (h1) in Tab.~\ref{tab:hfs}.
The value Borie gives on p.\,19 of Ref.\,\cite{Borie:2014:arxiv_v7} is
\begin{equation}
  \Delta E^\mathrm{B.}_\mathrm{Fermi}=6.14298\,\mathrm{meV}.
\end{equation}
It already includes the correction $\Delta E_\mathrm{\mu AMM}$ (h4)
due to the muon anomalous magnetic moment ($\mu$AMM).
However, Borie's value 
$\Delta E^\mathrm{B.}_\mathrm{Fermi}=6.14298\,\mathrm{meV}$
is not correct.
Since $\Delta E_\mathrm{Fermi}$ depends only on fundamental 
constants~\cite{Mohr:2012:CODATA10}, we have recalculated it
following Eq.\,(\ref{eq:EFermi}), 
and obtain a value of 
\begin{equation}
  \label{eq:ourEFermi}
  \Delta E_\mathrm{Fermi}= 6.14308\,\mathrm{meV}
\end{equation}
which differs from Borie's, but coincides with Martynenko's value\\
\begin{equation}
  \begin{aligned}
  \Delta E^\mathrm{M.}_\mathrm{Fermi} & = 6.1431\,\mathrm{meV} \\
  & = 6.1359\,\mathrm{meV}_\mathrm{h1} + 0.0072\,\mathrm{meV}_\mathrm{h4}.
  \end{aligned}
\end{equation}
Here, the two terms in the second line are, respectively, the 
Fermi splitting excluding the contribution 
of the anomalous magnetic moment of the muon, (h1), and 
the $\mu$AMM correction, (h4), which Martynenko calculates separately.

The Breit term $\Delta E_\mathrm{Breit}$ (h2) corrects for relativistic and 
binding effects accounted for in the Dirac-Coulomb wave function but 
excluded in the Schr\"odinger wave function. Both, Martynenko and Borie, 
calculated the Breit correction term to be
\begin{equation}
  \Delta E_\mathrm{Breit}=0.0007\,\mathrm{meV}.
\end{equation}

\subsection{Vacuum polarization (VP) and self-energy (SE) contributions}
\label{sec:HFS:QED}
On p.~21 of Ref.~\cite{Borie:2014:arxiv_v7}, Borie provides the values for $\epsilon_\mathrm{VP1}$ (h8) and
$\epsilon_\mathrm{VP2}$ (h5). The values we used are the ones for a point-like
nucleus. In Sec.\,\ref{sec:HFS:Rz} we give a correction due to the
finite size, which is given by (h25, h26). The corresponding terms from
Martynenko and Borie agree.

(h7) is neglected by Martynenko as pointed out on p.\,21 in
\cite{Borie:2014:arxiv_v7}.
The origin of the difference between Martynenko and Borie in (h9) is not
clear. In this case we take the average.
(h9b) is a correction in third order perturbation theory which is only 
given by Martynenko. 

The $\mu$VP contribution $\Delta E_{\mu \mathrm{VP}}$ (h12) is given
by Martynenko as
\begin{equation}
  \Delta E_{\mu \mathrm{VP}}=0.0002\,\mathrm{meV}.
\end{equation}
Borie included this contribution in the vertex term 
$\Delta E_\mathrm{vertex}$ (h13) as pointed out on p.\,21 in
\cite{Borie:2014:arxiv_v7}. As we do not use Borie's vertex term, we have
to consider the $\mu$VP term from Martynenko.

Martynenko gives a value for a term called 
{\em radiative nuclear finite size correction} (h17b). 
It is composed of four terms, 
{\em $\mu$SE with nuclear structure}, {\em jellyfish correction} 
and two vertex correction terms, so it
should also include (h13). Their sum
yields $-0.0005$\,meV. We think that (h14) is an additional term, only
calculated by Borie. So we add this to 'our choice'.

The hadron VP $\Delta E_\mathrm{hVP}$ (h18) results equal for both, Martynenko and Borie:
\begin{equation}
  \Delta E^\mathrm{M.,B.}_\mathrm{hVP} = 0.0002\,\mathrm{meV}.
\end{equation}
There is no considerable contribution from weak interaction \cite{Eides:2012:Weak}.

\subsection{Zemach radius}
\label{sec:HFS:Rz}
The Bohr-Weisskopf effect~\cite{BohrWeisskopf:1950} 
is the main finite size correction
to the 2S hyperfine splitting. It is also called
the Zemach term~\cite{Zemach:1956}, \EHFSZemach{},
and is listed as item (h20) in our summary.
The Zemach term is usually parameterized as~\cite{FriarSick:2004:Zemach}
\begin{equation}
\label{eq:HFS_Zemach}
\EHFSZemach{} = -\Delta E_\mathrm{Fermi}~2(Z\alpha)m_r~r_Z
\end{equation}
using the so-called Zemach radius of the nucleus~\cite{FriarSick:2004:Zemach}
\begin{equation}
\label{eq:Rz}
\begin{aligned}
r_Z & = \int d^3r \int d^3r'~r' \rho_E(\boldsymbol r)\rho_M(\boldsymbol r-\boldsymbol r') \\[1ex]
    & = - \frac{4}{\pi} \int_0^\infty \frac{dq}{q^2} \left( G_E(q^2) G_M(q^2) - 1 \right).
\end{aligned}
\end{equation}
This convolution of charge $\rho_E(\boldsymbol r)$ and magnetization
$\rho_M(\boldsymbol r)$ distribution comes from the fact, that the finite
charge distribution alters the muon's wave function at the origin.

Diagrammatically, the Zemach correction (h20) to the HFS, \EHFSZemach{}, 
is the elastic part of the two-photon exchange contribution to the
2S HFS, 
just like the Friar correction to the Lamb shift Fig.~\ref{fig:tpe}(a,b)
\cite{Pachucki:1996:LSmup}.\\
A recoil correction to the elastic TPE, (h23), is considered here, too.
It is somewhat parallel to the item \#22 of the Lamb shift, termed 
{\em rel.\ RC $(Z\alpha)^5$} listed in Tab.~\ref{tab:LS:QED}.\\
The inelastic part of the TPE correction to the 2S HFS, \EHFSinelast{}, 
is topic of the next Sec.~\ref{sec:HFS:Pol}.

Borie gives a value of
$\epsilon_\mathrm{Zemach}=-0.007398\,\mathrm{fm^{-1}}~r_Z$
(Ref.~\cite{Borie:2014:arxiv_v7}, p.\,22). 
Her Zemach contribution is hence (Eq.~(\ref{eq:HFS_Zemach}), 
Ref.~\cite{Borie:2014:arxiv_v7}, p.\,23 top)
\begin{equation}
  \EHFSZemach\,(\mathrm{Borie}) ~ = -0.04545 ~ r_Z ~ \mathrm{meV/fm}.
\end{equation}
Note that the coefficient 
\begin{equation}
-0.04545\,=\,\frac{3}{2}\,\beta_D\,(1+a_\mu)\,\epsilon_\mathrm{Zemach}
\end{equation}
does explicitly include the factor $(1+a_\mu)$.\\
Using $r_Z=2.593\pm0.016\,$fm from Ref.~\cite{FriarSick:2004:Zemach}
Borie's value for the Zemach contribution to the 2S HFS amounts to
\begin{equation}
  \EHFSZemach\,(\mathrm{Borie}) ~ = -0.11782\,\pm\,0.00074\,\mathrm{meV}.
\end{equation}
Borie mentions that nuclear recoil corrections (our item (h23)) are important,
but have not been included (Ref~\cite{Borie:2014:arxiv_v7}, p.\,22, bottom).

%


Martynenko calculates the 
{\em nuclear structure correction $\alpha^5$}
from the deuteron electromagnetic current that involves the form factors
$F_1$, $F_2$, and $F_3$ which can be related to the measured 
charge, magnetic and quadrupole form factors of the deuteron, 
$G_E(q^2)$, $G_M(q^2)$ and $G_Q(q^2)$,
see Eq.~(37) in Ref.~\cite{Faustov:2014:HFS_mud}.
Using the parameterization of $G_E$, $G_M$ and $G_Q$
from Ref.~\cite{Abbott:2000:EPJA7}, Martynenko obtains a
value of $-0.1163\pm0.0010$\,meV 
(Ref.~\cite{Faustov:2014:HFS_mud}, Tab.~1 item \#7, Eq.~(46)).\\
This value is the sum of the Zemach term, item (h20), as calculated by Borie,
but includes recoil corrections to the finite size effect,
(h23)~\cite{Martynenko:PC:2015}.
The separation of these two contributions is 
not unique~\cite{Martynenko:PC:2015}, but if one adopts the
canonical definition of the Zemach radius in terms of the form factors
$G_E$ and $G_M$ as given in Eq.~(\ref{eq:Rz}), 
one can separate Martynenko's sum into
\begin{widetext}
\begin{equation}
  \begin{aligned}
    \label{eq:MartZemRC}
    {\EHFSZemach}_\mathrm{+RC}\,(\mathrm{Martynenko}) 
    & = ~ \qquad \quad \quad -0.1163\,\mathrm{meV} & \pm ~ 0.0010\,\mathrm{meV}~\\
    & = ( -0.1178\,\mathrm{meV}_\mathrm{h20} + 0.0015\,\mathrm{meV}_\mathrm{h23}) & \pm ~ 0.0010\,\mathrm{meV}.
  \end{aligned}
\end{equation}
\end{widetext}
Here the item (h20) was calculated using the Zemach radius $r_Z = 2.5959$\,fm~\cite{Martynenko:PC:2015},
obtained by numerical integration of the parameterization of the deuteron form factors 
from Ref.~\cite{Abbott:2000:EPJA7}.
This Zemach radius is in excellent agreement
with the value $r_Z=2.593\pm0.016\,$fm from Ref.~\cite{FriarSick:2004:Zemach} 
used by Borie~\cite{Borie:2014:arxiv_v7}.
A small difference arises from
Martynenko's observation that the factor $(1+a_\mu)$ in 
Eq.~(\ref{eq:EFermi}) should {\em not} be included
for the $2\gamma$ amplitudes with point vertices~\cite{Martynenko:PC:2015}.\\
Rewriting the $-0.1178$\,meV of (h20) as
\begin{equation}
  \label{eq:h20}
  \EHFSZemach\,(\mathrm{Martynenko}) ~ = -0.0453934 ~ r_Z ~ \mathrm{meV/fm}
\end{equation}
makes the dependence on the Zemach radius explicit.
Putting the nuclear recoil corrections back in, the combined 
Zemach (h20) and recoil (h23)
corrections evaluated by Martynenko (Eq.~(\ref{eq:MartZemRC})) become
\begin{widetext}
\begin{equation}
  \begin{aligned}
    \label{eq:h20h23}
    {\EHFSZemach}_\mathrm{+RC} ~ 
    = -0.0453934 ~ r_Z ~ \mathrm{meV/fm}_\mathrm{h20}
    ~ + ~ (0.0015~\pm ~  0.0007)\,\mathrm{meV}_\mathrm{h23} 
  \end{aligned}
\end{equation}
\end{widetext}
which we adopt.
The total uncertainty of 0.0010\,meV given by Martynenko is then
the sum of the uncertainty in the Zemach
radius $\delta r_Z = 0.016$\,fm, corresponding to 0.0007\,meV,
and the uncertainty of (h23) given above.

\subsection{Nuclear polarizability contributions to the 2S HFS}
\label{sec:HFS:Pol}
The polarizability contribution to the 2S hyperfine splitting in 
muonic deuterium, \EHFSTPE{}, has only recently
been calculated for the first time by the group of 
Martynenko~\cite{Faustov:2014:HFS_mud}.
They obtain the polarizability term in two parts:\\ 
First, the {\em deuteron polarizability contribution} \EHFSTPE(deuteron)
(h22a)
is obtained from the analytic expressions derived in zero range approximation 
for electronic deuterium by Khriplovich and 
Milstein~\cite{Khriplovich:2004:inelastic}.
This part takes into account the virtual excitation of a deuteron made 
from point nucleons.

Second, the much smaller {\em internal deuteron polarizability contribution} 
$\Delta E_\mathrm{int.~d-pol.}$ (h22b), which accounts for the excitation of the
individual nucleons (proton and neutron) inside the deuteron. This part
is estimated based on the results for muonic 
hydrogen~\cite{Faustov:2003:PRA67_052506}.

Summing these two up yields
\begin{widetext}
\begin{equation}
  \label{eq:HFS:pol}
  \Delta E^\mathrm{M.}_\mathrm{tot.~d-pol.} = 0.2121(42)\,\mathrm{meV} + 0.0105(25)\,\mathrm{meV}=0.2226(49)\,\mathrm{meV}
\end{equation}
\end{widetext}
with a generous uncertainty that accounts also for the fact that the original
derivation~\cite{Khriplovich:2004:inelastic} was for electronic, 
and not for muonic deuterium.

Eq.~(\ref{eq:HFS:pol}) is also the value quoted by Borie. As pointed out by 
Borie, Ref.~\cite{Borie:2014:arxiv_v7} p.\,22,
it is not clear whether the 'elastic' contribution of the two-photon exchange
diagrams is taken into account.

As for the Lamb shift, the polarizability term for the 2S HFS 
is the one with by far the largest uncertainty.

\subsection{Further corrections to the 2S HFS}
\label{sec:HFS:other}
Several further corrections are considered by either Martynenko or Borie 
(h24, h25, h26, h27, and h27b):\\
Martynenko calculates a mixed term which includes eVP
and a nuclear structure correction (h24) 
\begin{equation}
  \Delta E^\mathrm{M.}_\mathrm{eVP+nucl. struct.}=0.0019\,\mathrm{meV}
\end{equation}
as well as two further nuclear structure contributions (h27, h27b):
\begin{equation}
  \Delta E^\mathrm{M.}_\mathrm{nucl. str. corr.}=0.0008\,\mathrm{meV}
\end{equation}
\begin{equation}
  \Delta E^\mathrm{M.}_\mathrm{nucl. str. SOPT}=-0.0069\,\mathrm{meV}
\end{equation}
Borie gives finite size corrections (h25, h26) to the eVP terms $\epsilon_\mathrm{VP1}$ and
$\epsilon_\mathrm{VP2}$, both contributing -0.00068\,meV. These are obtained
by calculating the difference of $\epsilon_\mathrm{VP1}$ and
$\epsilon_\mathrm{VP2}$ with a point size nucleus compared to the ones when
considering a finite size (\cite{Borie:2014:arxiv_v7}, p.\,21).
It is not yet clear whether the nuclear structure contributions from Martynenko
are complementary to the ones of Borie. We should remark that item (h27b) is 
quite big. It doesn't seem to be included in Borie's calculations and is the
main reason for the difference between Borie~\cite{Borie:2014:arxiv_v7}
and Martynenko~\cite{Faustov:2014:HFS_mud}.\\
For now we refrain from assigning a large uncertainty to this item h27b,
but an independent calculation, or at least an estimate of it's accuracy, 
would certainly be helpful.

\subsection{Total 2S hyperfine splitting}
Hence, collecting all terms, but separating out the deuteron polarizability correction 
Eq.~(\ref{eq:HFS:pol}) as it is the dominant source of uncertainty, we
can write the total 2S HFS in muonic deuterium as
\begin{widetext}
\begin{equation}
  \begin{aligned}
    \Delta E_\mathrm{HFS}^\mathrm{th}
    & = 6.17415(73)\,\mathrm{meV}  + 0.22260(490)\,\mathrm{meV} & - ~ 0.04539\,r_Z\,\mathrm{meV}&\\
    & = \qquad \qquad   \HFSVALERR\,\mathrm{meV}              & - ~ 0.04539\,r_Z\,\mathrm{meV}&.
  \end{aligned}
\end{equation}
\end{widetext}
The large uncertainty in the polarizability corrections to the 2S HFS
will prevent a determination of the deuteron Zemach radius from the
measured transitions in muonic deuterium~\cite{CREMA:muD}.
An improved calculation of the polarizability terms is therefore highly desirable.

Using the Zemach radius $r_Z=(2.593 \pm 0.016)\,\mathrm{fm}$ \cite{FriarSick:2004:Zemach} we get:
\begin{equation}
  \Delta E_\mathrm{HFS}^\mathrm{th}=6.27905(495)\,\mathrm{meV}
\end{equation}
to be compared to the muonic deuterium measurement~\cite{CREMA:muD}.
Alternatively, one can use the measurement and the Zemach radius to accurately determine
the polarizability contributions. Such a number may serve as a benchmark for accurate
lattice calculations.

\onecolumngrid
\clearpage
\begin{table}[h!]
  \onecolumngrid
  \footnotesize
  \setlength\extrarowheight{2pt}
  \centering
  \caption{All contributions to the {\bf 2S hyperfine splitting (HFS)} in 
    muonic deuterium.
    The item numbers h$i$ in the first column follow
    the entries in Tab.~3 of Ref.~\cite{Antognini:2013:Annals}.
    For Martynenko, numbers \#1 to \#15 refer to rows in 
    Tab.\,I of Ref.\,\cite{Faustov:2014:HFS_mud}, 
    whereas numbers in parentheses refer to equations therein.
    Borie~\cite{Borie:2014:arxiv_v7} gives the values as {\em coefficients}
    to be multiplied with the sum of (h1+h4). We list the resulting values 
    in meV.
    AMM: anomalous magnetic moment,
    PT: perturbation theory,
    VP: vacuum polarization,
    SOPT: second order perturbation theory,
    TOPT: third order perturbation theory.\newline
    All values are in meV (meV/fm for h20).}
    \label{tab:hfs}
    \hspace*{-7mm}\begin{tabular}{l|l|f{5} l l| f{5} l l| f{5} l l}
      \hline
      \hline
      & \cnt{1}{Contribution} 
                    & \cnt{3}{Martynenko \cite{Faustov:2014:HFS_mud}} 
                                       & \cnt{3}{Borie \cite{Borie:2014:arxiv_v7}} 
                                                         & \cnt{3}{Our choice}\\
      \hline
      h1& Fermi splitting, $(Z\alpha)^4$          
                    & 6.1359 &&\#1, (6)
                                       &          &&              
                                                         &         &  &\\
      h4& $\mu$AMM corr., $\alpha(Z\alpha)^4$  
                    & 0.0072 &&\#2, (7)
                                       &          &&
                                                         &         &  &\\
      sum& (h1+h4)                 
                    & 6.1431 &&               
                                       & 6.14298  &&p.\,19
                                                         & 6.14308 
                                                                   & 
                                                                      & Eq.~(\ref{eq:ourEFermi})\\
      h2& Breit corr., $(Z\alpha)^6$           
                    & 0.0007 &&\#3, (8)
                                       & 0.00069  &&p.\,19
                                                         & 0.00069 &  & B\\
      \hline
      h5& eVP in 2nd-order PT, $\alpha(Z\alpha)^5$ ($\epsilon_\mathrm{VP2}$)
                    & 0.0207 &&\#4, (23) 
                                       & 0.02070  &&p.\,21         
                                                         & 0.0207  &  & M\\
      h7& Two-loop corr.\ to Fermi-energy ($\epsilon_\mathrm{VP2}$)
                    & \lft{2}{neglected} &             
                                       & 0.00016  &&p.\,21         
                                                         & 0.00016 &  & B\\
      h8& One-loop eVP in $1\gamma$ int., $\alpha(Z\alpha)^4$ ($\epsilon_\mathrm{VP1}$)
                    & 0.0134 &&\#4, (12)
                                       & 0.01339  &&p.\,21         
                                                         & 0.0134  &  & M\\
      h9& Two-loop eVP in $1\gamma$ int., $\alpha^2(Z\alpha)^4$ ($\epsilon_\mathrm{VP1}$)
                    & 0.0005 &&\#5, (16), (29-32)
                                       & 0.00010  &&p.\,21         
                                                         & 0.0003  &$\pm\,0.0002$  
                                                                      & avg.\\
      h9b& VP contr. in TOPT
                    & 0.00004 &&\#6, (33)
                                       &          &&             
                                                         & 0.00004 &  & M\\
      \hline
      h12& $\mu$VP (sim.\ to $\epsilon_\mathrm{VP}$)
                    &0.0002  &&\#9, (48)
                                       & \lft{2}{incl.\ in h13} &p.\,21 
                                                         & 0.0002  &  & M\\
      \hline
      h13& Vertex, $\alpha(Z\alpha)^5$          
                    & \lft{2}{incl.\ in h17b} &         
                                       &-0.00059  &&p.\,19         
                                                         & \lft{2}{incl.\ in h17b}
                                                                      & \\
      h14& Higher order corr.\ of (h13), part with ln($\alpha$)
                    &        &&              
                                       & -0.00004 &&p.\,19         
                                                         &-0.00004 &  & B\\
      h17b& Radiative nucl.\ fin.\ size corr., $\alpha(Z\alpha)^5$
                    & -0.0005 &&\#13, (71-74)
                                       &          &&             
                                                         & -0.0005 &  & M\\
      \hline
      h18& Hadron VP, $\alpha^6$               
                    & 0.0002  &&\#10, (50)
                                       & 0.00016  &&p.\,19         
                                                         & 0.00016 &  & B\\
      h19& Weak interact.\ contr.               
                    & 0       && p.\,10            
                                       & 0
                                                  &&p.\,21         
                                                         & 0       &  & \cite{Eides:2012:Weak}\\
      \hline
      h20& Fin.\ size (Zemach) corr.\ to $\Delta E_\mathrm{Fermi}$, $(Z\alpha)^5$
                    & -0.04539\,r_Z
                              &
                               & \cite{Martynenko:PC:2015}~\footnote{The published value for the sum of items h20+h23 is $-0.1163 \pm 0.0010$\,meV \cite{Faustov:2014:HFS_mud}. For the separation into items h20 and h23 see text.}
                                       & -0.04545~r_Z\footnote{Calculated from Eq.~(\ref{eq:EFermi}), including the
factor $(1+a_\mu)$. According to Martynenko, this factor should be omitted in the $2\gamma$ amplitudes with point vertices.}
                                                  & 
                                                   &p.\,22         
                                                         & 0.04539~r_Z &
                                                                      & M\\
      h23& Recoil corr.\ to fin.\ size
                    & 0.0015
                               &$\pm 0.0007$
                                 & \cite{Martynenko:PC:2015}~\footnotemark[1]
                                       & 
                                                  &&      
                                                         & 0.0015  &$\pm 0.0007$
                                                                      & M\\
      sum & (h20+h23) 
                    & -0.1163
                              &$\pm\,0.0010$
                               &\#7, (46)
                                       &
                                                  & 
                                                   &
                                                         & 
                                                                   & 
                                                                      & M\\

      \hline
      h22a& Deuteron polarizability, $(Z\alpha)^5$
                    & 0.2121 &$\pm\,0.0042$  
                              &\#14 using \cite{Khriplovich:2004:inelastic}    
                                       &          &&            
                                                         & 0.2121  &$\pm\,0.0042$
                                                                      & M\\
      h22b& Deuteron internal polarizability, $(Z\alpha)^5$
                    & 0.0105 &$\pm\,0.0025$  
                              &\#15 using \cite{Faustov:2003:PRA67_052506}
                                       &          && 
                                                         & 0.0105  &$\pm\,0.0025$
                                                                      & M\\
      sum& (h22a+h22b)
                    & 0.2226 &$\pm\,0.0049$
                              &
                                       & 0.2226   &$\pm\,0.0049$ 
                                                   &p.\,22        
                                                         &         &  &\\
      \hline
      h24& eVP + nucl.\ struct.\ corr., $\alpha^6$
                    & 0.0019 &$\pm\,0.00001$ 
                              &\#8, (47)
                                       &          &&            
                                                         & 0.0019  & 
                                                                      & M\\
      h25& eVP corr.\ to fin.\ size (sim.\ to $\epsilon_\mathrm{VP2}$)
                    &        &&
                                       & -0.00068  && p.\,21      
                                                         & -0.00068
      \footnote{Difference of two terms in Borie~\cite{Borie:2014:arxiv_v7}. See text.}
                                                                   &  & B\\
      h26& eVP corr.\ to fin.\ size (sim.\ to $\epsilon_\mathrm{VP1}$)
                    &        &&
                                       & -0.00068  && p.\,21      
                                                         &-0.00068 &  & B\\
      h27& Nucl.\ struct.\ corr., $\alpha(Z\alpha)^5$
                    & 0.0008 &&\#11, (55)
                                       &           &&            
                                                         & 0.0008  &  & M\\
      h27b&Nucl.\ struct.\ in SOPT               
                    & -0.0069 &&\#12, (59)
                                       &           &&            
                                                         & -0.0069 &  & M\\
      \hline
      \hline
      &&&&&&&&&&\\
      & \bf Sum
                    & 6.39824 &$\pm\,0.00494$
                               &
                                       &  6.39880  &$\pm\,0.00490$
                                                    &            
                                                         & \TABHFSVAL & \bf $\pm$\,\TABHFSERR
                                                                      &\\
      &
                    & -0.04539\,r_Z
                              &&
                                       &  -0.04545\,r_Z
                                                   &&
                                                         & \bf -0.\bf 04539\,r_Z
                                                                    &
                                                                      &\\
      &&&&&&&&&&\\
      \hline
      \hline
    \end{tabular}
\end{table}

\clearpage


\begin{table}[h]
  \onecolumngrid
  \footnotesize
  \setlength\extrarowheight{1pt}
  \caption{
    Contributions to the 2P fine structure. The items (f7a), (f7d), and (f7e)
    originate from the same graphs as the Lamb shift items \#11, \#12,
    and \#30$^*$,
    respectively.
    VP: vacuum polarization, AMM: anomalous magnetic moment, KS: K\"all\'en-Sabry.\newline
    All values are in meV.}
  \label{tab:fs}
  \begin{tabular}{l|l  |f{7}         |f{7}          |f{7} l    |f{6} l l}
    \hline
    \hline
       & Contribution  & \cntl{1}{Martynenko}
                                     & \cntl{1}{Borie}
                                                    & \cntl{2}{Karshenboim}
                                                                 & \cnt{3}{Our choice} \\
       &               & \cntl{1}{\cite{Martynenko:2014:muD_Theory} Tab.2} 
                                     & \cntl{1}{\cite{Borie:2014:arxiv_v7}} 
                                                    &  \cntl{2}{~} 
                                                                 &     &
                                                                                  &     \\
    \hline
    f1 & Dirac
                       &             &  8.86430     &            &&     &         &     \\
    f2 & Recoil        
                       &             & -0.02521     &            &&     &         &     \\
    f3 & Contrib.\ of order $(Z\alpha)^4$
                       &  8.83848    &              &            &&     &         &     \\
    f4 & Contrib.\ of order $(Z\alpha)^6$ and $(Z\alpha)^6\,m_1/m_2$
                       &  0.00030    &              &            &&     &         &     \\
    sum&  (f1+f2) or (f3+f4)
                       &  8.83878    & 8.83909      &            &
                                                                 & 8.83894       & $\pm\,0.00016$ & avg. \\
    \hline
    f5 & eVP correction (Uehling), $\alpha(Z\alpha)^4$  
                       &  0.00575    &  0.00575     & 0.0057361                   & \cite{Karshenboim:2012:PRA85_032509} Tab.IV
                                                                 &  0.0057361     &         & K   \\
    f6 & 2nd order eVP corr.\ (KS), $\alpha^2(Z\alpha)^4$
                       &  0.00005    &  0.00005     &  0.0000501                  & \cite{Korzinin:2013:PRD88_125019} Tab.IX ``eVP2''
                                                                 &  0.0000501     &         & K   \\
    f7a& $\alpha^2(Z\alpha)^4 m$, like \#11
                       &             &              &  0.0000127                  & \cite{Korzinin:2013:PRD88_125019} Tab.IX (a)
                                                                 &  0.0000127     &         & K   \\
    f7d& $\alpha^2(Z\alpha)^4 m$, like \#12
                       &             &              &  0.0000991                  & \cite{Korzinin:2013:PRD88_125019} Tab.IX (d)
                                                                 &  0.0000991     &         & K   \\
    f7e& $\alpha^2(Z\alpha)^4 m$, like \#30$^*$
                       &             &              &  0.0000012                  & \cite{Korzinin:2013:PRD88_125019} Tab.IX (e)
                                                                 &  0.0000012     &         & K   \\
    \hline
    f8 & AMM (second order)
                       &             &  0.01949     &            &&            
                                                                        &         &     \\
    f9 & AMM (higher orders)
                       &             &  0.00007     &            &&            
                                                                        &         &     \\
    sum& Total AMM (f8+f9)
                       &  0.01957    &  0.01956     &            && 0.019565 
                                                                        &$\pm\,0.000005$&avg. \\
    \hline
    f10 & Finite size
                       & -0.00028    & -0.00027     &            && -0.000274 
                                                                        & &\footnote{This is item (r8),
 evaluated for the deuteron radius from Eq.~(\ref{eq:Rd_muH}), see text.} \\
    \hline
    \hline
    &&&&&\\
       & \bf Sum           &  8.86387    &  8.86419     &            &
                                                                 &  \TABFSVAL & \bf $\pm$\,\TABFSERR &  \\
    &&&&&\\
    \hline
    \hline
  \end{tabular}
\end{table}
\twocolumngrid

\section{2P Fine Structure}
\label{sec:fs}

The contributions to the 2P$_{3/2}-$2P$_{1/2}$ fine structure splitting 
in muonic deuterium are displayed in Tab.~\ref{tab:fs}.

The main contributions to the fine structure have only been 
calculated by Borie~\cite{Borie:2014:arxiv_v7} and Martynenko's 
group~\cite{Krutov:2011:PRA84_052514,Martynenko:2014:muD_Theory}.
For the latter, we refer to the more recent paper 
Ref.~\cite{Martynenko:2014:muD_Theory}. The values agree in both papers.

As always, Borie starts from the Dirac equation, which has to be corrected 
for recoil effects. This sum of entries (f1)+(f2) has to be compared to
Martynenko's leading term (f3), corrected for relativistic effects (f4) which
are automatically included in the Dirac equation.
The result of both approaches agree reasonably well and we adopt the average
$8.83894 \pm\,0.00016$\,meV.

The relativistic recoil correction of order $\alpha(Z\alpha)^4$, (f5), 
has been calculated including all recoil corrections of order $m/M$ 
by Karshenboim~\cite{Karshenboim:2012:PRA85_032509}. This value
thus supersedes~\cite{Karshenboim:PC:2015}
the values obtained by Borie~\cite{Borie:2014:arxiv_v7}
and Martynenko~\cite{Krutov:2011:PRA84_052514,Martynenko:2014:muD_Theory}.

The K\"all\'en-Sabry term, item (f6), agrees nicely among all authors.
Karshenboim \etal{} have evaluated some higher order 
$\alpha^2(Z\alpha)^4 m$ contributions with great
accuracy~\cite{Korzinin:2013:PRD88_125019} which we list as (f7a), (f7d), and (f7e).
These terms originate from the same graphs as the Lamb shift items \#11, \#12,
and \#30$^*$,
shown in Figs.~\ref{fig:item_11}, \ref{fig:item_12} and \ref{fig:item_30}, respectively.

Contributions from the anomalous magnetic moment of the muon in second (f8) and 
higher (f9) orders have been calculated by Borie and Martynenko~\etal{}, and their sums
agree.

A finite size correction to the $2P_{1/2}$ state (f10) is included, too. This term
is the same term as item (r8) of the \rr-dependent contributions to the Lamb shift,
but with the opposite sign
(see discussion in Sec.~\ref{sec:LS:Radius}).
We evaluate (r8) for the deuteron radius from Eq.~(\ref{eq:Rd_muH}) and
obtain for our choice of the item (f10)
$-0.000606\,\rr{} = -0.000274 \pm 6\cdot10^{-8}$\,meV.

Summing up we obtain as our choice for the
2P$_{3/2}-$2P$_{1/2}$ fine structure splitting 
in muonic deuterium
\begin{equation}
  \label{eq:fs}
  \Delta E_\mathrm{fs}(\mathrm{2P}_{3/2}-\mathrm{2P}_{1/2}) = \FSVALERR{}\,\mathrm{meV}.
\end{equation}
%

\section{2P Levels}
\label{sec:2PHFS}

The various 2P levels displayed in Fig.~\ref{fig:energy_level} are separated
by the 2P fine structure treated in Sec.~\ref{sec:fs}, 
and further split by the 2P hyperfine splitting caused by the 
magnetic hyperfine interaction and the
electric quadrupole interaction.
The Breit-Pauli Hamiltonian can be displayed in matrix form 
as a sum of the magnetic HFS matrix and the quadrupole interaction matrix:
\begin{widetext}
  \begin{equation}
    \label{eq:2P_matrix}
    \begin{aligned}
M_{\rm Breit-Pauli} ~
  &= &  &
    \begin{pmatrix}
     -1.381777 &   0        &  -0.126405 &   0        &   0        \\
      0        &   0.690889 &   0        &  -0.199864 &   0        \\
     -0.126405 &   0        &   8.161148 &   0        &   0        \\
      0        &  -0.199864 &   0        &   8.582931 &   0        \\
      0        &   0        &   0        &   0        &   9.285903
    \end{pmatrix}_{\rm magnetic~HFS}\\[2ex]
  &~ &+ &
   \begin{pmatrix}
      0        &   0        &  0.613872 &   0        &   0        \\
      0        &   0        &  0        &  -0.194123 &   0        \\
    ~~0.613872 &   0        &\,0.434073 &   0        &   0        \\
      0        &\,-0.194123 &  0        &\,-0.347258 &   0        \\
      0        &   0        &  0        &   0        & \,0.086815
    \end{pmatrix}_{\rm quadrupole~int.}
   \\[3ex]
  &= & & \bordermatrix{
              &2P_{1/2}^{F=1/2}&2P_{1/2}^{F=3/2}&2P_{3/2}^{F=1/2}&2P_{3/2}^{F=3/2}&2P_{3/2}^{F=5/2}\\[1ex]
2P_{1/2}^{F=1/2} & -1.381777   &  0          &  0.487467    &  0          &  0          \\[1ex]
2P_{1/2}^{F=3/2} &  0          &  0.690889   &  0           & -0.393988   &  0          \\[1ex]
2P_{3/2}^{F=1/2} &  0.487467   &  0          &  8.595220    &  0          &  0          \\[1ex]
2P_{3/2}^{F=3/2} &  0          & -0.393988   &  0           &  8.235672   &  0          \\[1ex]
2P_{3/2}^{F=5/2} &  0          &  0          &  0           &  0          &  9.372718   \\[1ex]
} ~ \mathrm{meV.}
    \end{aligned}
  \end{equation}
\end{widetext}

\begin{table}[b!]
  \setlength\extrarowheight{3pt}
  \setlength\tabcolsep{1ex}
  \centering
  \caption{Elements of the transition matrix (Eq.\,(\ref{eq:2P_matrix}))
    from Ref.~\cite{Borie:2014:arxiv_v7}.
    See also Ref.~\cite{Brodsky:1967:zeemanspectrum}.
    The various quantities are explained in the text.
  }
  \label{tab:mat_elements}
  \begin{tabular}{l l l}
    \cnt{1}{$j$}  & \cnt{1}{$j'$} & \cnt{1}{Energy}  \\ 
    \hline
    \cnt{3}{magnetic HFS}\\
    \hline
    1/2  & 1/2  & $(\beta_D'/6)(2+x_\mathrm{d}+a_\mu)[-\delta_{F,1/2}+1/2\,\delta_{F,3/2}]$  \\[1ex]
    3/2  & 3/2  & $\Delta E_\mathrm{fs} ~+~ (\beta_D'/4)(4+5\,x_\mathrm{d}-a_\mu)\times$ \\[1ex]
         &      & ~ \qquad $[-1/6\,\delta_{F,1/2}-1/15\,\delta_{F,3/2}+1/10\,\delta_{F,5/2}]$\\[1ex]
    3/2  & 1/2  & $(\beta_D'/48)(1+2\,x_\mathrm{d}-a_\mu)[-\sqrt{2}\,\delta_{F,1/2}-\sqrt{5}\,\delta_{F,3/2}]$  \\[1ex]
    \hline
    \cnt{3}{quadrupole interaction}\\
    \hline
    1/2  & 1/2  & 0  \\[1ex]
    3/2  & 3/2  & $\epsilon_Q ~ [\delta_{F,1/2}-4/5\,\delta_{F,3/2}+1/5\,\delta_{F,5/2}]$  \\[1ex]
    3/2  & 1/2  & $\epsilon_Q ~ [\sqrt{2}\,\delta_{F,1/2}-1/\sqrt{5}\,\delta_{F,3/2}]$  \\[1ex]
    \hline
  \end{tabular}
\end{table}
It attains off-diagonal elements
from mixing of levels with the same total angular momentum $F$, but different
total muon angular momentum 
$j$~\cite{Brodsky:1967:zeemanspectrum,Pachucki:1996:LSmup,Jentschura:2011:AnnPhys1,Borie:2014:arxiv_v7},
as shown in Tab.~\ref{tab:mat_elements}. 
Note that the diagonal terms of the
quadrupole interaction lead to a change in the order 
of the 2P$_{3/2}$ levels (see also Fig.\,\ref{fig:energy_level}).

We follow Borie's treatment~\cite{Borie:2014:arxiv_v7}, 
see also Pachucki~\cite{Pachucki:1996:LSmup}
and Jentschura~\cite{Jentschura:2011:AnnPhys1},
but use our value for the 2P fine structure 
$\Delta E_\mathrm{fs}(2P_{3/2}-2P_{1/2}) = \FSVALERR{}\,\mathrm{meV}$
from Sec.~\ref{sec:fs}, Eq.~(\ref{eq:fs}), as well as
a more recent value of the deuteron quadrupole moment
\begin{equation}
  \label{Qdeut}
  Q ~ = ~ 0.285783\,(30)~\mathrm{fm}^2
\end{equation}
from Ref.~\cite{Pavanello:2010:d_quadmom}, or, equivalently, 
$Q=7.33945(77) \cdot 10^{-24}$/meV$^2$,
using $\hbar c = 197.3269718(44)$\,MeV\,fm~\cite{Mohr:2012:CODATA10}.

\begin{table}[t]
  \setlength\extrarowheight{4pt}
  \setlength\tabcolsep{2ex}
  \centering
  \caption{Input parameters for the transition matrix. We recalculated Borie's 
    values, but use our fine structure (see Sec.\,\ref{sec:fs}) and an updated 
    value of the quadrupole moment Q \cite{Pavanello:2010:d_quadmom} for 
    $\epsilon_Q$, see Eq.\,(\ref{eq:quadmom}).}
  \label{tab:inputparameter}
  \begin{tabular}{l f{9} f{13}}

                   & \lft{1}{Borie \cite{Borie:2014:arxiv_v7}} 
                                & \lft{1}{our value}  \\[0.5ex] 
    \hline
    $\beta_D$          &  4.0906\,\textrm{meV}  &  4.0906259\,\textrm{meV} \\
    $\beta_D'$         &  4.0922\,\textrm{meV}  &  4.0922253\,\textrm{meV} \\
    $\epsilon_Q$       &  0.43439\,\textrm{meV} &  0.434073(46)\,\textrm{meV}        \\
    $x_\mathrm{d}$      &  0.0248                &  0.0247889 \\[1mm]
    a$_\mu$            &  \multicolumn{2}{c}{0.00116592 \qquad \qquad}\\[1mm]
    $\Delta E_\mathrm{fs}$  
                       &  8.86419\,\textrm{meV} &  \FSVALERR\,\textrm{meV} \\
    \hline
  \end{tabular}
\end{table}

The numerical values of the quantities used in Tab.~\ref{tab:mat_elements} 
are given in Tab.~\ref{tab:inputparameter}. 
In brief, $\beta_D$ is defined in Eq.~(\ref{eq:betaD}). For the 2P levels,
\begin{equation}
  \beta_D' = \beta_D (1 + \epsilon_{2P})
\end{equation}
has to be used which contains the Uehling correction required for 
levels with $\ell > 0$ (see p.~25 and Eq.~(12) in Ref.~\cite{Borie:2014:arxiv_v7}).
For muonic deuterium,
\begin{equation}
  \label{eq:eps2p}
  \epsilon_{2P}=0.000391.
\end{equation}

The quadrupole moment of the deuteron
enters in the hyperfine splitting via the quadrupole interaction, 
see Borie~\cite{Borie:2014:arxiv_v7}, pp.\ 24 and 25.
\begin{equation}\label{eq:quadmom}
  \epsilon_Q = \frac{\alpha Q}{2} \frac{(\alpha Z m_r)^3}{24} (1+\epsilon_{2P}),
\end{equation}
where 
$Q$ the quadrupole moment of the deuteron
and 
$\epsilon_{2P}$ is given in Eq.~(\ref{eq:eps2p}).

\begin{table}[b!]
  \setlength\extrarowheight{7pt}
  \centering
  \caption{2P levels from fine- and hyperfine splitting.
    All values are in meV relative to the 2P$_{1/2}$ level.
    The fine structure (2P$_{3/2} - $2P$_{1/2}$ energy splitting)
    is our value 
    $\Delta E_\mathrm{fs}=\FSVALERR\,$meV from Eq.\,(\ref{eq:fs}).
    %
    %
    Uncertainties arise from the quadrupole moment $Q$ in Eq.~(\ref{Qdeut})
    and 
    $\Delta E_\mathrm{fs}$.
    }
  \label{tab:hfsP}
  \begin{tabular}{l f{7} f{8}}
                   & \lft{1}{Borie \cite{Borie:2014:arxiv_v7}} 
                                & \lft{1}{Our value}    \\[0.5ex]
    \hline
    $2P_{1/2}^{F=1/2}$ & -1.4056   & -1.40554(1)           \\
    $2P_{1/2}^{F=3/2}$ &  0.6703   &  0.67037(1)           \\
    \hline
    $2P_{3/2}^{F=1/2}$ &  8.6194   &  8.61898(17)          \\
    $2P_{3/2}^{F=3/2}$ &  8.2560   &  8.25619(16)          \\
    $2P_{3/2}^{F=5/2}$ &  9.3728   &  9.37272(16)          \\
    \hline
  \end{tabular}
\end{table}

Diagonalizing the matrix Eq.~(\ref{eq:2P_matrix})
results in shifts of the 2P$(F=1/2)$ and 2P$(F=3/2)$ levels by 
\begin{equation}
  \begin{aligned}
    \label{eq:DeltaMix}
    \Delta_{1/2} & = 0.02376\,\mathrm{meV~and}\\
    \Delta_{3/2} & = 0.02052\,\mathrm{meV},
  \end{aligned}
\end{equation}
respectively,
as displayed in Fig.~\ref{fig:energy_level}.
The resulting energies of the various 2P sublevels are summarized in 
Tab.~\ref{tab:hfsP}.

\section{Summary}
In summary, we have compiled all known contributions to the Lamb shift, 
the 2P fine structure,
 and the 2S and 2P hyperfine splittings,
from QED and nuclear structure contributions.

For the Lamb shift, the QED contributions in Tab.~\ref{tab:LS:QED}
show good agreement between the four (groups of) authors. A problem with
our item \#2 from Ref.~\cite{Krutov:2011:PRA84_052514} was identified and
resolved by the auhors. Ultimately, the uncertainty of these
 ``pure QED'' terms in Tab.~\ref{tab:LS:QED} is sufficiently good.\\
For the radius-dependent terms in Tab.~\ref{tab:LS:Radius} we find good
agreement between the authors, too. Some terms have however been calculated by
only one group. We recalculate a small term (r3') to verify that the 
model-dependence imposed by this contribution is sufficiently small.\\
The main limitation for the Lamb shift, and hence the deuteron charge radius
to be extracted from the upcoming data, originates from the two-photon exchange
contribution to the Lamb shift in \mud. Here, a superficial inspection of
the six modern values published in five papers since 2011~\cite{Pachucki:2011:PRL106_193007,Friar:2013:PRC88_034004,Carlson:2014:PRA89_022504,Hernandez:2014:PLB736_344,Pachucki:2015:PRA91_040503} vary between 1.68\,meV and 2.01\,meV,
suggesting an uncertainty as large as 0.3\,meV.\\
The term-by-term comparison of the individual contributions in 
Tab.~\ref{tab:LS:Pol} revealed that the agreement is in fact much better.
Fruitful discussions with the authors of these papers and other experts 
in the field revealed missing terms and resulted in updated values
of some individual terms.
It is very reassuring that vastly different approaches give results in
excellent agreement, when corrected for missing terms:
zero-range approximation,
modern nuclear models like AV18 (from two groups of authors) and 
$\chi$EFT-inspired NN-forces up to N$^3$LO order,
and dispersion relations using electron-deuteron scattering data.
Our average, $\POLVALFINAL \pm \POLERRFINAL$\,meV
is a reliable prediction for the deuteron polarizability contribution
to the Lamb shift in \mud.

For the 2S-HFS, several nuclear structure contributions have so far only been 
calculated by one group~\cite{Faustov:2014:HFS_mud}:
These are items (h22a), (h22b), and (h27b) in Tab.~\ref{tab:hfs},
which are rather large, and their uncertainties dominate the theoretical
uncertainty for the 2S-HFS. This uncertainty will prevent us from
obtaining a meaningful value of the Zemach radius of the deuteron from
the measurement of the 2S-HFS in \mud.
An improved calculation of these items is therefore desirable.

For the 2P fine- and hyperfine splittings we collect all terms from the 
various authors, recalculate the matrix elements of the Breit-Pauli
Hamiltonian with updated values of the 2P fine structure and the 
deuteron quadrupole moment.
Diagonalizing this matrix Eq.~(\ref{eq:2P_matrix}) we obtain the 2P level 
energies, and their uncertainties.

\section{Acknowledgements}
\label{sec:ack}

We we are grateful to 
S.~Bacca, 
M.~Birse,
E.~Borie, 
M.~Eides,
J.L.~Friar,
M.~Gorchtein,
A.P.~Martynenko,
J.~McGovern,
K.~Pachucki, and
M.~Vanderhaeghen
for their careful reading of earlier versions of the manuscript and their
very valuable remarks and insightful discussions in particular of the
issues concerning the nuclear polarizability contributions to the Lamb shift 
discussed in Sec.~\ref{sec:LS:Pol}.
S.~Karshenboim is acknowledged for helpful remarks on QED theory in general.

\appendix
\section{The Darwin-Foldy term}
\label{app:DF}

The Darwin-Foldy term, which is part of the Barker-Glover corrections
(our item~\#17 in Tab.~\ref{tab:LS:QED}) has historically been subject
of different definitions.

Pachucki and Karshenboim~\cite{PachuckiKarshenboim:1995} argue that the DF
term originates from the Zitterbewegung of the nucleus and is hence
absent for a spin-1 nucleus such as the deuteron 
(as well as for the spin-0 $^4$He nucleus).

Khriplovich, Milstein and Sen'kov~\cite{Khriplovich:1996:DF} argue that 
the DF term must be made a part of the rms charge radius to be consistent 
with electron scattering. In this case, the DF term is {\em not} absent for
spin-1 nuclei such as the deuteron.

Friar, Martorell and Sprung~\cite{Friar:1997:PRA56_4579} have emphasized 
that the DF-term can
be alternatively considered as part of a recoil correction of order $1/M^2$,
or as the energy shift due to a part of the mean-square radius of the nuclear
charge distribution.
They advocate the second choice but admit that the first choice has to be
used for the proton because "it is unfortunately far too late to change these
conventions for the hydrogen atom". They recommend, however, to not extend the
hydrogen atom conventions to other nuclei.

Jentschura has discussed the situation in some
breadth~\cite{Jentschura:2011:DF} and concluded that the DF term should indeed
be considered a contribution to the atomic energy levels due to the nuclear
Zitterbewegung, supporting the ``atomic physics'' convention of
Ref.~\cite{PachuckiKarshenboim:1995}. The DF term is hence absent for the 
deuteron.

This ``atomic physics'' convention, in which the DF term is {\em not} a part
of the rms charge radius, but rather a recoil correction of order 
$(Z\alpha)^4m^3/M^2$ to
the energy levels, is the convention used in
CODATA-2010~\cite{Mohr:2012:CODATA10}, see Eq.~(26) and (27) therein.  It is
also the convention used in the most recent measurement of the H-D isotope
shift~\cite{Parthey:2010:PRL_IsoShift,Jentschura:2011:IsoShift} which is the
origin of the difference of the squared rms radii of the deuteron and the
proton given in Eq.~(\ref{eq:iso_HD}). Moreover, it is the convention used
for the proton radius in muonic hydrogen~\cite{Antognini:2013:Annals}.

Therefore, to be able to directly compare the numerical values of the proton
and deuteron rms charge radii obtained in electronic and muonic atoms, one
must follow the ``atomic physics'' convention~\cite{PachuckiKarshenboim:1995,
Jentschura:2011:IsoShift,Jentschura:2011:DF,Mohr:2012:CODATA10}, which is what we do.

\section*{References}

\bibliography{ref}

\begin{thebibliography}{82}
\bibitem [{\citenamefont {{CREMA Collaboration}}()}]{CREMA:muD}%
  \BibitemOpen
  \bibfield  {author} {\bibinfo {author} {\bibnamefont {{CREMA
  Collaboration}}},\ }\href@noop {} {}\bibinfo {note} {To be
  published.}\BibitemShut {Stop}%
\end{thebibliography}

\end{document}